\documentclass[conference]{IEEEtran}
\IEEEoverridecommandlockouts

\usepackage{cite}
\usepackage{amsmath,amssymb,amsfonts}
\usepackage{graphicx}
\usepackage{textcomp}
\usepackage{kotex}
\usepackage{url}
\usepackage{subcaption}
\usepackage{comment}
\usepackage{booktabs}
\usepackage{float}
\usepackage{tabularx}
\usepackage{hyperref}
\usepackage{tikz}
\usepackage{amsmath,amssymb,amsfonts}
\usepackage{textcomp}
\usepackage[normalem]{ulem}
\hypersetup{hidelinks}
\usepackage{wrapfig}

\usepackage{adjustbox}
\usepackage{algorithm}
\usepackage{algpseudocode}
\usepackage{etoolbox}

\usepackage{multirow}
\usepackage[margin=1in]{geometry}
\usepackage{makecell}  
\usepackage{enumitem}
\usepackage{setspace}
\usepackage{microtype}

\let\oldbibliography\thebibliography
\renewcommand{\thebibliography}[1]{%
  \oldbibliography{#1}%
  \setlength{\itemsep}{2.95pt}%
  \setlength{\baselineskip}{7pt}
  \setlength{\lineskiplimit}{-\maxdimen}
}

\newcommand*\bcircled[1]{\tikz[baseline=(char.base)]{
    \node[shape=circle,draw,fill=black,text=white,inner sep=0.5pt, line width=0.5pt] (char) {\bfseries\scriptsize #1};}}
\newcommand*\wcircled[1]{\tikz[baseline=(char.base)]{
    \node[shape=circle,draw,fill=white,text=black,inner sep=0.5pt, line width=1pt] (char) {\bfseries\scriptsize #1};}}

\usepackage{minted}
\newcommand{\proposed}{\textsc{CALL}}
\newcommand{\costlru}{\textsc{CLRU}}
\newcommand{\windowlru}{\textsc{WLRU}}
\newcommand{\fifo}{\textsc{FIFO}}

\newcommand{\squishlist}{
\begin{list}{$\bullet$}
	{ \setlength{\itemsep}{0pt}      \setlength{\parsep}{-0pt}
		\setlength{\topsep}{4pt}       \setlength{\partopsep}{0pt}
		\setlength{\listparindent}{-2pt}
		\setlength{\itemindent}{-5pt}
		\setlength{\leftmargin}{1em} \setlength{\labelwidth}{0em}
		\setlength{\labelsep}{0.5em} } }
\newcommand{\squishend}{
\end{list}  }

\def\BibTeX{{\rm B\kern-.05em{\sc i\kern-.025em b}\kern-.08em
    T\kern-.1667em\lower.7ex\hbox{E}\kern-.125emX}}

\begin{document}

\title{CALL: Context-Aware Low-Latency Retrieval in Disk-Based Vector Databases}

\author{\IEEEauthorblockN{Yeonwoo Jeong$^{1}$, Hyunji Cho$^{1}$, Kyuli Park$^{1}$, Youngjae Kim$^{1}$, Sungyong Park$^{1}$\IEEEauthorrefmark{2}
\thanks{\IEEEauthorrefmark{2}S. Park is the corresponding author.
}
}
\IEEEauthorblockA{$^{1}$\textit{Sogang University},
Seoul, Republic of  Korea \\
\{akssus12, llawliet37, kyuripark, youkim, parksy\}@sogang.ac.kr
}}

\maketitle

\begin{abstract}
Embedding models capture both semantic and syntactic structures of queries, often mapping different queries to similar regions in vector space.
This results in non-uniform cluster access patterns in modern disk-based vector databases.
While existing approaches optimize individual queries, they overlook the impact of cluster access patterns, failing to account for the locality effects of queries that access similar clusters.
This oversight increases cache miss penalty.
To minimize the cache miss penalty, we propose \proposed{}, a context-aware query grouping mechanism that organizes queries based on shared cluster access patterns.
Additionally, \proposed{} incorporates a group-aware prefetching method to minimize cache misses during transitions between query groups and latency-aware cluster loading.
Experimental results show that \proposed{} reduces the 99th percentile tail latency by up to 33\% while consistently maintaining a higher cache hit ratio, substantially reducing search latency.
\end{abstract}


\vspace{-6pt}
\section{Introduction}
\begin{figure}[!t]
	\centering
    \includegraphics[width=0.8\linewidth]{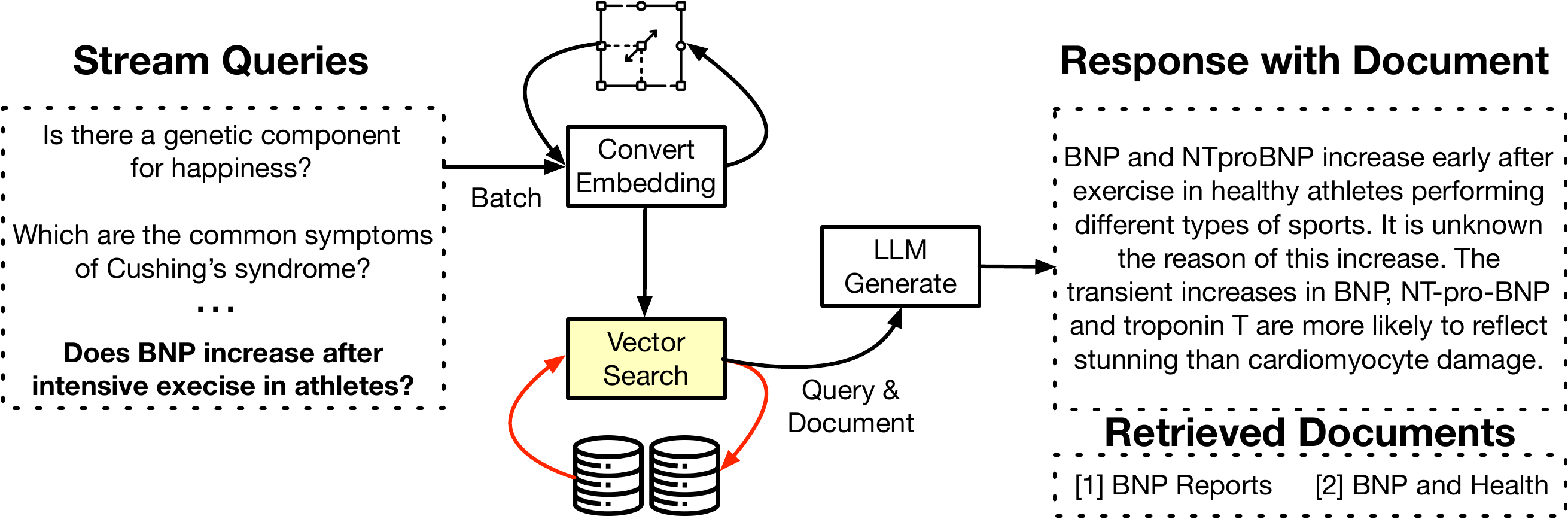}
    \caption{A workflow of stream-based RAG application. \textcolor{red}{Red} arrows indicate vector search in the RAG pipeline.}
    \label{fig:streamrag}
    \vspace{-25pt}
\end{figure}

Large Language Models (LLMs) often generate responses that include incorrect information, when prompted with queries beyond their training data~\cite{hallucination}.
To mitigate this limitation, Retrieval-Augmented Generation (RAG) applications enhance user prompts by integrating relevant domain-specific documents retrieved from a vector database~\cite{ragsystem5}.
In these systems, unstructured documents are embedded into high-dimensional vectors and indexed, allowing the prompt's embedding to be compared against them using similarity metrics (e.g., inner product or cosine similarity) for grounded response generation.
Recently, RAG applications are increasingly deployed as backend cloud services, handling continuous streams of user queries in diverse domains such as real-time clinical guidance~\cite{healthrag} and interactive assistant agents~\cite{chatbotllm}.
These \textit{stream-based RAG applications} frequently encounter concurrent queries, as illustrated in Figure~\ref{fig:streamrag}, which require low latency for each response.
In stream-based RAG applications, timely document retrieval is crucial for supporting fast LLM inference, making low-latency vector search a critical challenge.

In practice, vector databases commonly load entire vector index into memory. 
While the in-memory vector search offers high-speed retrieval, the size of vector indexes often exceeds available memory capacity~\cite{diskann, starling, edgerag}.
For example, constructing one billion floating-point vectors in 96 dimensions requires over 350 GB of memory, which often surpasses the memory capacity of a typical single server~\cite{starling}.
To address this memory constraint, two alternative solutions have been explored: distributed vector search and disk-based vector search.
Distributed vector search~\cite{milvus} partitions the index across multiple servers and parallelizes the search, allowing the system to scale with increasing index size.
While effective, this approach incurs significant infrastructure costs, as larger indexes require additional memory-heavy index servers.
In contrast, disk-based vector search~\cite{diskann, starling, edgerag} stores the entire vector index on high-speed secondary storage such as NVMe SSDs.
This approach eliminates the need for memory-intensive servers and offers a cost-effective alternative by leveraging a single machine equipped with fast storage.
These systems partition the vector index into multiple clusters and retrieve only the relevant clusters on demand at runtime.

Although partial clusters are loaded into memory at runtime, I/O remains a key bottleneck in disk-based vector databases~\cite{starling}.
Empirical studies reveal that I/O accounts for over 90\% of the total search latency, while computation constitutes less than 10\%~\cite{diskann}.
To minimize I/O overhead, modern vector databases incorporate application-level cache mechanisms~\cite{gptcache, edgerag, diskann}.
For instance, FAISS~\cite{faiss}, a widely used vector database, employs a cluster cache to retain frequently accessed cluster files, which are partitioned subsets of the full vector index.
Before accessing storage, FAISS checks the cluster cache. If the required cluster is cached, it avoids disk access, otherwise it retrieves the cluster from the disk, incurring latency.
To mitigate cache miss penalties, prior works have proposed several techniques such as frequency-based caching~\cite{gptcache, edgerag}.
However, these approaches often overlook deeper system-level bottlenecks that arise under multi-user query workloads.

Our preliminary studies using the RAG benchmark datasets~\cite{beir} reveal that query access patterns exhibit non-uniform cluster locality. (Section~\ref{subsec:motivation}). 
In stream-based RAG applications, where queries are processed in batches, this non-uniform locality presents a valuable opportunity to optimize cluster cache utilization.
However, prior work generally treats incoming queries as independent, missing opportunities for scheduling across batched queries. 
In contrast, our work focuses on the streaming scenario, where batch-level query scheduling can significantly reduce overall latency. 

We identify three key problems that limit the performance benefits of in-batch query scheduling.

\noindent\textbf{Non-Uniform Cluster Access Patterns.} 
Modern embedding models encode text into dense vector representations that capture both semantic meaning and syntactic structures (e.g., question formats and specific phrasings).
As a result, even queries with different content within the same batch may be mapped to similar regions in the vector space.
However, existing approaches process incoming queries in order and fail to consider contextual similarities among queries within the batch.
This lack of awareness leads to suboptimal cache utilization. 
Reordering queries that access a high proportion of the same cluster files can significantly improve cluster locality in vector databases. 

\noindent\textbf{Prefetch Invisibility.} 
While such grouping increases cache hit rates within each group, it introduces a new challenge at group boundaries.
Transitions between groups with disjoint cluster access patterns can lead to abrupt cache evictions.
Because the vector database is unaware of the group structure, it cannot prefetch the necessary clusters in advance.
As a result, some clusters are evicted prematurely, reducing cache efficiency and increasing tail latency.

\noindent\textbf{Imbalanced Cluster Loading.} 
Even with well-formed query groups and effective prefetching, some clusters inevitably need to be fetched from disk at runtime.
This introduces a system-level challenge that can reduce parallel efficiency during vector search.
Cluster files vary significantly in size due to uneven document distribution or inconsistent vector density across shards~\cite{imbalance_kmeans}. 
Modern vector search engines such as Milvus~\cite{milvus} and Qdrant~\cite{qdrant} typically assign homogeneous tasks (e.g., loading documents or generating embedding) to a shared job queue processed in parallel by multiple worker threads. 
Similarly, cluster loading tasks enter a shared queue and are dispatched to threads in round-robin order. 
When large cluster files are added to the queue without considering their size, they become stragglers, occupying a thread for a long time and stalling parallel I/O progress, which causes severe load imbalance among worker threads.

\noindent\textbf{Our Approach.} 
This paper proposes \proposed{}, \underline{c}ontext-\underline{a}ware \underline{l}ow \underline{l}atency retrieval in disk-based vector databases,
which jointly addresses the three aforementioned problems: non-uniform cluster access patterns, prefetch invisibility, and imbalanced cluster loading.

First, we propose a context-aware grouping method that analyzes cluster access patterns of batched queries and partitions them into virtual groups based on similarity. 
While content or keyword-based query grouping has proven effective in web caching and recommendation systems, these approaches are fundamentally unsuitable for vector databases.
This is mainly because queries in vector databases are represented as dense, high-dimensional embeddings, which inherently lack explicit semantic tokens or features suitable for keyword-based analysis.
Furthermore, relying solely on vector similarity does not guarantee optimal caching behavior, as semantically similar queries may still access disjoint sets of clusters.
Instead, our approach shifts caching granularity from individual queries to groups of queries sharing similar cluster access patterns.

Second, 
we introduce a group transition-hinted prefetching 
algorithm, which uses group-level metadata to prefetch clusters ahead of group transitions, reducing cache miss penalties.
It attaches lightweight metadata that marks group boundaries, enabling the system to anticipate cache needs and reduce cache evictions during group switches.

Finally, 
we present a load-weighted greedy packing 
algorithm which prioritizes large cluster files and distributes them evenly across available worker threads.
This approach mitigates the long-tail task execution that can reduce overall efficiency in parallel cluster loading.

In summary, our key contributions are as follows:
\squishlist
\item{We reveal that 
queries processed in batches 
exhibit non-uniform cluster access patterns, offering opportunities to improve cache utilization through query reordering.}
\item{We identify why disk-based vector databases using cluster caches fail to improve search performance despite employing cache replacement policies.}
\item{
We propose \proposed{}, a system designed to integrate seamlessly with various caching mechanisms employed in modern vector databases.}
\item{Our extensive evaluation across various datasets demonstrates that \proposed{} outperforms baseline methods by up to 33\% in 99th percentile tail latency and achieves up to an 84\% improvement in end-to-end search latency, all without introducing additional overhead for grouping.}
\squishend
\vspace{-6pt}
\section{Background and Related Works}
\label{sec:background}

\subsection{Vector Database}
\label{sec:bg_rag}
Vector databases are specialized systems designed to store, index, and search large-scale, high-dimensional embeddings.
The core functionality of a vector database consists of two main stages: indexing and retrieval.

\noindent\textbf{Indexing.} 
Rather than storing all vectors in a flat array, Approximate Nearest Neighbor (ANN) indexing techniques organize vector embeddings to prune the search space. 
ANN methods are typically categorized into graph-based and clustering-based algorithms.


\begin{itemize}
    \item Graph-based indexing builds a navigable proximity graph over the vector space.
    Each node (e.g., vector) is connected to a subset of other nodes based on local similarity, forming a small-world topology.
    The graph encodes neighborhood relationships that can be exploited for efficient traversal during search. 
    This offers high search accuracy and fast convergence but typically requires high memory overhead.
    \item Clustering-based indexing partitions the vector space into disjoint regions, using unsupervised clustering algorithms like k-means~\cite{kmeans}. 
    This allows the search to focus on a small number of candidate regions, reducing computation and disk I/O.
    This approach is particularly effective for disk-based deployments due to its compact index representation.
    In this paper, we employ the clustering-based indexing method to implement disk-based ANN vector search.
\end{itemize}

\noindent\textbf{Retrieval.} Once the index has been constructed, vector search is performed by encoding the user query into a dense embedding using the same model used during indexing.
The query vector is traversed through the index to find a subset of candidate vectors likely to be similar.
Similarity scores between the query vector and the candidate vectors are computed using distance metrics such as Euclidean distance. 
The system then returns the top-k vectors with the highest similarity scores.
\vspace{-0.01in}

\subsection{Disk-based ANN Vector Search}
\label{sec:bg_ann}
\vspace{-1pt}

A disk-based ANN vector search mechanism is proposed to address the memory limitations of in-memory vector search, where the entire index often cannot fit into DRAM. 
To overcome this, this technique offloads large-scale vector index to high-speed storage and fetches on-demand indexes needed for ANN search~\cite{diskann, starling}. 
Disk-based ANN vector search commonly adopts clustering-based indexing algorithms~\cite{edgerag, faiss}.

\begin{figure}[!t]
    \centering
    \setlength{\tabcolsep}{0.5em} 
    \renewcommand{\arraystretch}{1.0} 
	\begin{tabular}{@{}c@{}c@{}c@{}}
            \includegraphics[width=0.3\linewidth]{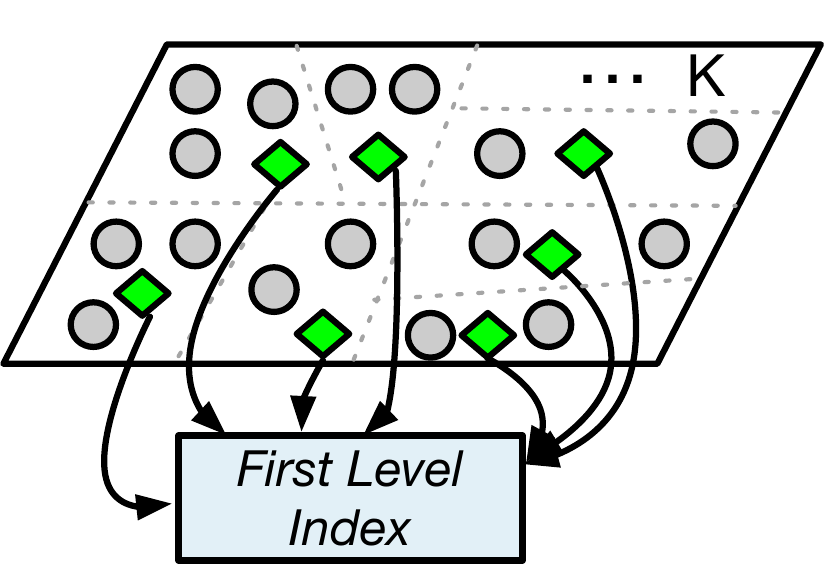} &
        \includegraphics[width=0.5\linewidth, trim={0 0 0 0cm},clip]{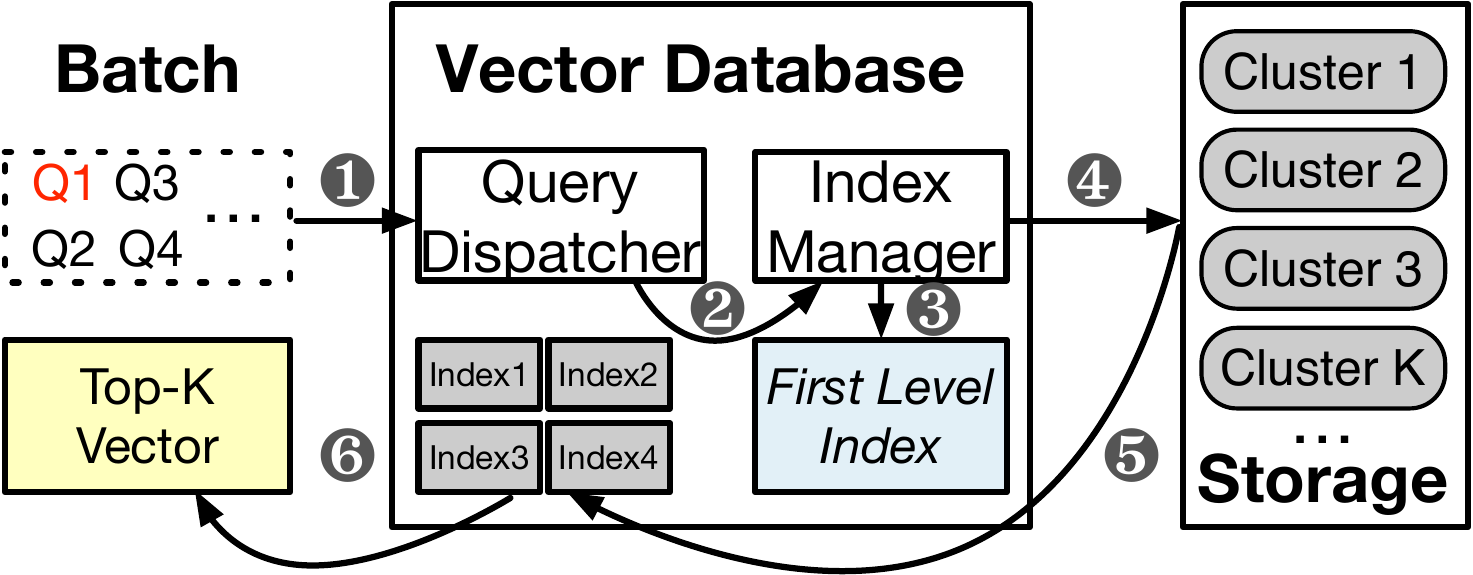} \\
        \small (a) IVF Indexing &
        \small (b) Disk-based vector search\\
    \end{tabular}
        \caption{IVF indexing and disk-based vector search workflow. In (a), circles in \textcolor{gray}{gray} represent vectors, while \textcolor{green}{green} diamonds denote centroids.
        }
    \label{fig:vectorsearch}
    \vspace{-20pt}
\end{figure}

Figure~\ref{fig:vectorsearch} (a) shows a workflow of Inverted File (IVF~\cite{ivf}) indexing, one of the widely used clustering-based algorithms. 
The entire vector space is first partitioned into k regions through k-means clustering.
During training, k-means iteratively refines a set of centroids by alternating between two steps: assigning each vector to its nearest centroid and updating each centroid as the mean of its assigned vectors. 
This process repeats until convergence, producing a set of representative centroids that partition the space into disjoint clusters.
K clusters with the disjoint vector space are stored on disk as vector arrays.
These centroids are stored in a first-level index, which resides entirely in memory. 
At query time, the vector database uses the first level index to perform an initial comparison, narrowing the search scope.

Figure~\ref{fig:vectorsearch} (b) shows a disk-based vector search using IVF indexing.
It follows a two-level search process.
Upon receiving batched queries (e.g., Q1, Q2, Q3, and Q4), each query is encoded into a multi-dimensional vector~\bcircled{\footnotesize{1}}. 
The query dispatcher sends the query vectors to index manager~\bcircled{\footnotesize{2}}.
Then, the index manager scans the first-level index to identify the centroids closest to each query vector~\bcircled{\footnotesize{3}}.
The vector database exposes a parameter, \textit{nprobe}, which controls the number of clusters scanned among the clusters.
In this example, four out of k clusters are selected for Q1's vector search.
Based on the found centroids, the index manager requests four clusters from the  storage~\bcircled{\footnotesize{4}} and constructs a partial index~\bcircled{\footnotesize{5}}.
The partial index, built from the selected clusters, enables top-k search by restricting computation to a small subset of the full vector index.
Finally, Q1 is searched over the indexes, after which the vector database ranks the results by distance, returning the k nearest vectors~\bcircled{\footnotesize{6}}.



\subsection{Related Works}
\label{sec:related}
\vspace{-1pt}
To date, there have been several works to enhance the performance in disk-based ANN vector search under constrained memory conditions. 
Most existing work focuses on optimizing cache utilization at the individual query level by applying traditional cache replacement policies~\cite{diskann, edgerag, gptcache}.


Caching Management adopts a multi-tiered storage architecture, where frequently accessed clusters are retained in memory, while infrequently accessed clusters are stored in storage.
DiskANN~\cite{diskann} introduces a disk-based ANN search algorithm optimized for large-scale vector datasets.
DiskANN keeps a partial search graph in memory and lazily loads additional nodes and edges from SSD.
To cut random I/O, it caches frequently accessed graph data for a subset of high-access vertices in DRAM.

EdgeRAG~\cite{edgerag} is a disk-based vector search framework that indexes only the clusters relevant to each query at runtime.
At the indexing stage, it profiles per-cluster embedding latency (e.g., disk reads and temporary index construction).
Using these metrics, it applies a cost-aware LRU policy that prioritizes clusters that are either frequently accessed or slow to load.

GPTCache~\cite{gptcache} presents an application-level semantic cache store for entire LLM responses.
GPTCache caches entire LLM responses to reuse prior results when semantically similar queries arrive.
It checks the embedding similarity between the incoming request and cached responses, returning the responses if the embedding similarity exceeds the predefined threshold.
Rather than caching full responses, our approach caches shared clusters, allowing partial indexes to be reused across semantically similar queries.

These works have aimed to improve cache efficiency at the individual query level, but they overlook the varying cluster access patterns among queries within a batch. 
Consequently, they fail to exploit inter-query cluster locality, leading to redundant cluster loads, poor cache utilization, and ultimately lower search performance in batched query scenarios.

\vspace{-3pt}
\section{Motivations}
\label{subsec:motivation}
\begin{figure}[!t]
    \centering
    \setlength{\tabcolsep}{0.5em} 
    \renewcommand{\arraystretch}{1.0} 
	\begin{tabular}{@{}c@{}c@{}c@{}}
            \includegraphics[width=0.45\linewidth]{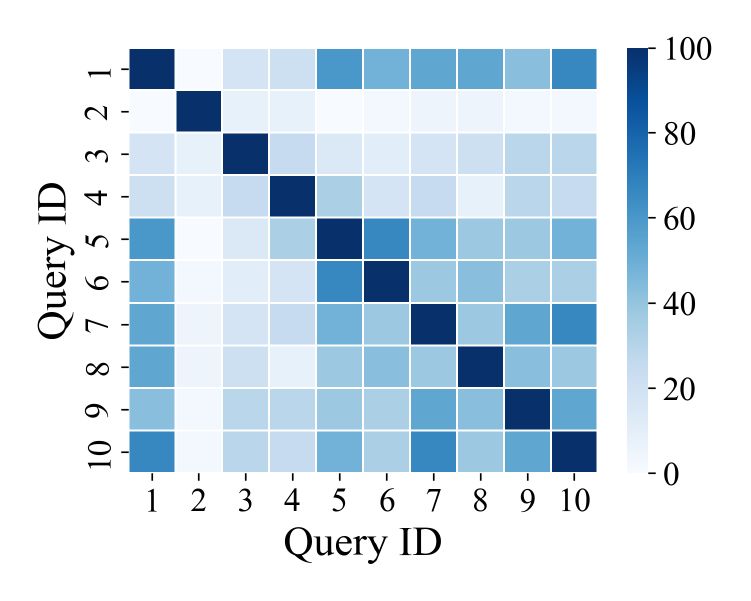} &
        \includegraphics[width=0.45\linewidth, trim={0 0 0 0cm},clip]{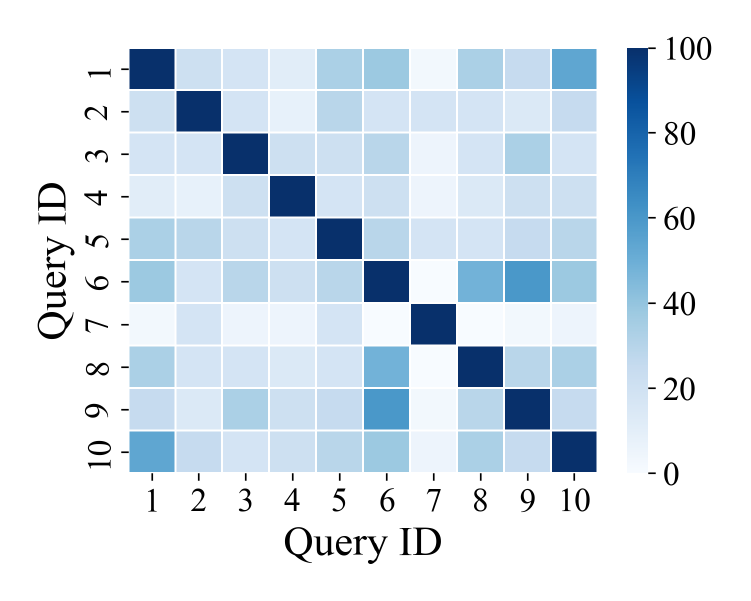} \\
        \small (a) all-miniLM-L6-v2~\cite{minilm} &
        \small (b) gte-modernbert-base~\cite{gte}\\
    \end{tabular}
        \caption{Cluster accessed patterns per embedding model.}
    \label{fig:moti1_non_uniform_pattern}
    \vspace{-16pt}
\end{figure}


\subsection{Non-Uniform Cluster Access Patterns}
\vspace{-1pt}
To analyze cluster access patterns, we conducted a synthetic experiment to measure the similarity in the batched queries.
For this, we used hotpotqa dataset~\cite{hotpotqa} from BEIR benchmark suite~\cite{beir}, using its query set as input.
HotpotQA is a large-scale multi-hop question answering dataset, widely used to evaluate the retrieval performance of LLM-based QA tasks, as each query requires combining evidence from multiple documents.
Since our system processes queries in batches, a burst of arrivals leads to larger batch sizes, which reflects practical scenarios where multiple queries arrive concurrently.
To enable query bursts occurring within short time windows, we generated query traffic following a Weibull distribution and selected FAISS~\cite{faiss} as the vector search engine.

We employ all-miniLM-L6-v2 and gte-modernbert-base to account for variation in embedding dimensionality and learning objectives, which are 384 and 768 dimensions, respectively.
MiniLM is trained via knowledge distillation for efficient general-purpose semantic matching, whereas GTE is instruction-tuned to better align with retrieval-oriented tasks.
This selection allows us to evaluate our system across models with different capacities, alignment strategies, and dimensionalities.

To quantify similarity, we use the Jaccard index~\cite{jaccard}, which measures the overlap between the set of all clusters accessed by quersies in a batch.
The Jaccard index computes the overlap between the sets of clusters per query and represents it as a ratio.
We set the total number of clusters to 100 and the query fan-out to 10, ensuring that each query accessed 10 clusters.

Figure~\ref{fig:moti1_non_uniform_pattern} depicts the cluster access patterns for each embedding model. 
The heatmaps show the degree of cluster overlap across a sequence of queries, where darker shades represent higher similarity.
We randomly sampled 10 queries from the evaluation set to construct the heatmap, where both the X and Y axes represent query indices in the sampled sequence (i.e., query execution order).
Notably, the similarity between adjacent queries is often low, while certain non-adjacent query pairs exhibit high overlap, indicating non-uniform cluster patterns.

For instance, Query 1 and 2 exhibit minimal overlap, whereas Query 1 and 10 share over 60\% of their accessed clusters.
Similarly, Figure~\ref{fig:moti1_non_uniform_pattern} (b) shows that Query 6, 8, and 9 exhibit an overlap, each sharing more than 50\% of their accessed clusters.
These observations stem from the behavior of the embedding model.
The embedding models capture both structural and semantic similarities when converting queries into vector representations, leading to diverse cluster distributions across the queries. 
Many queries follow similar syntactic patterns (e.g., "What year did Einstein win the Nobel Prize?" or "What year was Google founded?").
Since embedding models learn these recurring structures, they often map structurally similar queries to nearby regions in the vector space, even when their topics differ.
\textit{This finding implies that reordering batch queries based on dynamic cluster access patterns can significantly enhance cache utilization.}

\begin{figure}[!t]
    \centering
    \setlength{\tabcolsep}{0.5em} 
    \renewcommand{\arraystretch}{1.0} 
	\begin{tabular}{@{}c@{}c@{}c@{}}
            \includegraphics[width=0.5\linewidth]{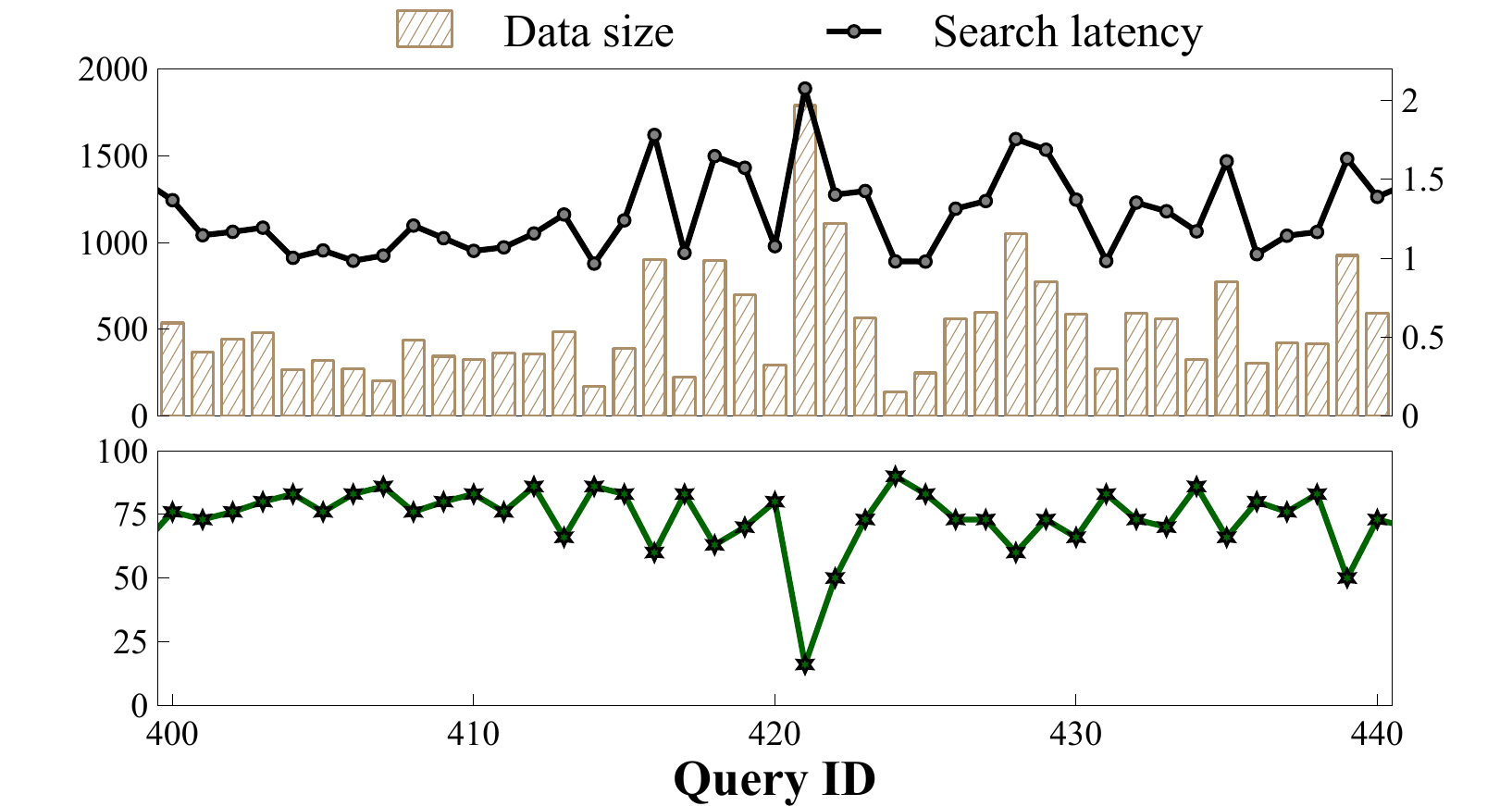} &
        \includegraphics[width=0.5\linewidth, trim={0 0 0 0cm},clip]{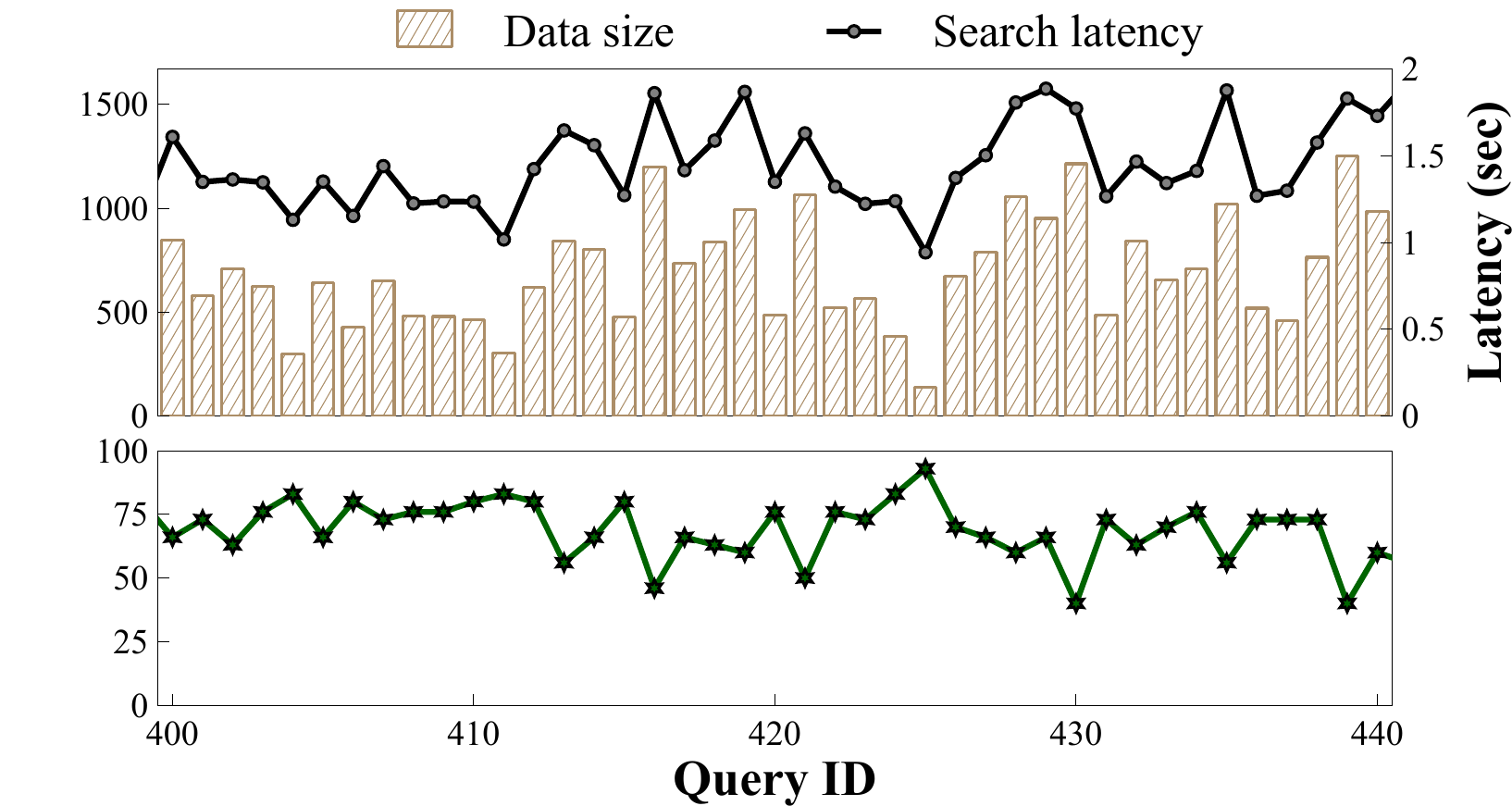} \\
        \small (a) \windowlru{} &
        \small (b) \fifo{}\\
    \end{tabular}
        \caption{Time-series cache hit ratio, cumulative read size of the clusters, and end-to-end search latency under two cache replacement policies.}
    \label{fig:moti2_cacheutil}
    \vspace{-16pt}
\end{figure}


\vspace{-6pt}
\subsection{Analysis of Replacement Effects on Cluster Cache}
\vspace{-1pt}
Cache replacement policies ignoring the non-uniform cluster patterns fall short of maximizing cache efficiency.
To verify this, we applied two cache replacement policies in the same scenario.
One is \windowlru{} that retains only the top 10 most frequently accessed clusters within each 1-minute window. 
The other is \fifo{}, which evicts the cluster that has remained in the cache the longest.

Figure~\ref{fig:moti2_cacheutil} presents cumulative read size of the clusters, and search latency per cache replacement policy based on the cache hit ratio.
Across all policies, we observe distinct performance characteristics.

As shown in Figure~\ref{fig:moti2_cacheutil} (a), \windowlru{} exhibits significant performance degradation in certain scenarios.
For example, at Query 421, the cache hit rate drops to 20\%. 
The query incurs a latency spike of 2.2 seconds while reading 1.8 GB of cluster files from storage, approximately 9 times higher than when the hit rate is 80\%.
This demonstrates the window boundary problem: clusters just outside the current window are evicted, even if they remain highly relevant to upcoming queries.
In addition, \fifo{} policy shows higher cache utilization fluctuations.
Cache hit ratios frequently drop into about 60\%, and the corresponding latencies exhibit more frequent and pronounced spikes, occasionally approaching 2 seconds. 
The results indicate that the existing caching policy fails to capture access patterns across queries, leading to unstable and highly fluctuated cache hit rates.

\begin{figure}[!t]
    \centering
    \setlength{\tabcolsep}{0.5em} 
    \renewcommand{\arraystretch}{1.0} 
	\begin{tabular}{@{}c@{}c@{}c@{}}
            \includegraphics[width=0.5\linewidth]{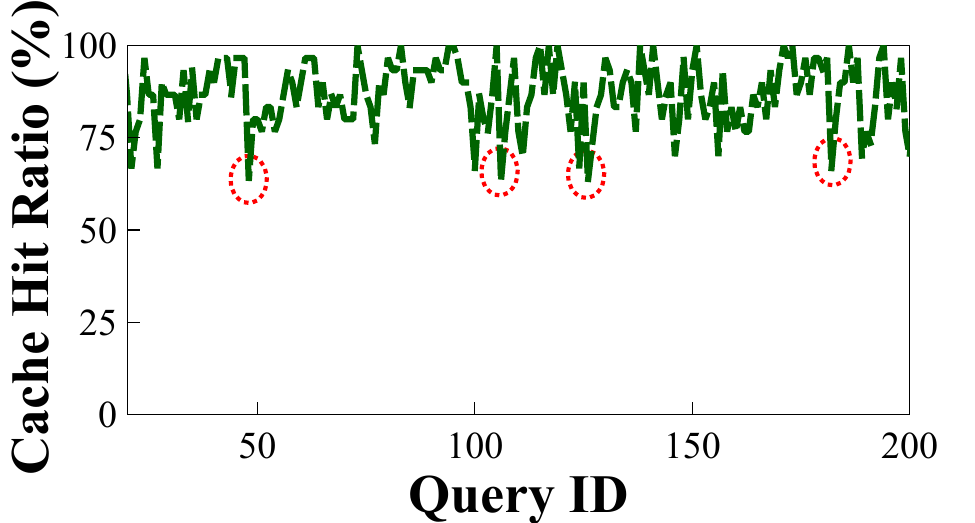} &
        \includegraphics[width=0.5\linewidth, trim={0 0 0 0cm},clip]{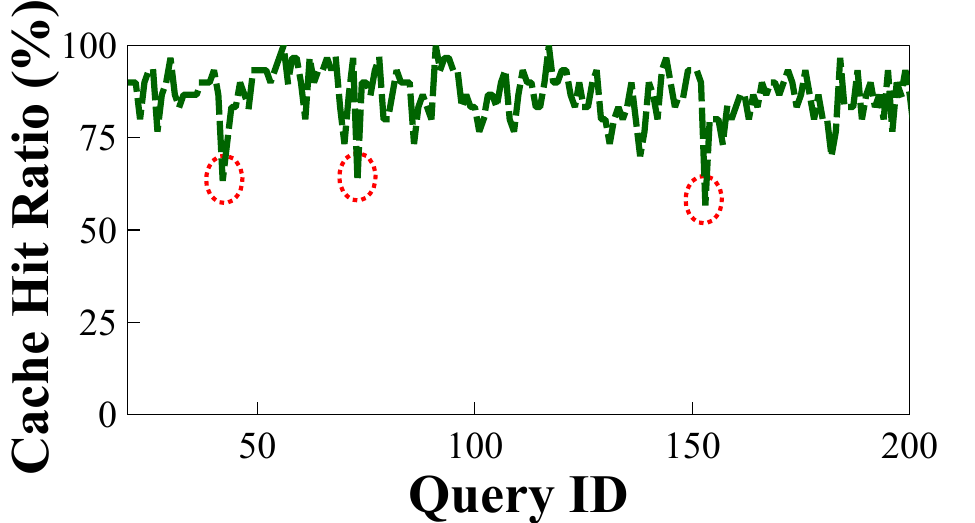} \\
        \small (a) Grouping + CLRU &
        \small (b) Grouping + WLRU \\
    \end{tabular}
        \caption{Time-series cache hit ratio trend applying query grouping and prefetch schemes. \textcolor{red}{Red} dotted circles indicate cache hit ratio drop by query group transition.}
    \label{fig:moti2_cache_drop}
    \vspace{-15pt}
\end{figure} 

\vspace{-6pt}
\subsection{Prefetch Invisibility} 

While grouping queries based on similar cluster access patterns appears to be a logical approach to enhance cache efficiency, it does not always yield higher cache hit rates.
To investigate the cause of inconsistent cache hit rates despite query grouping, we performed an experiment where queries were grouped based on Jaccard similarity.
Detail of our query grouping is described as Section~\ref{subsec:context_qg}.
Additionally, we extend both \costlru{} and \windowlru{} with a prefetching scheme.
For example, \costlru{} prefetches clusters with the highest loading costs every minute, while \windowlru{} prefetches the most frequently accessed clusters, each with a fixed prefetching degree of 20.
Despite these efforts to prefetch clusters, we consistently observed noticeable drops in cache hit rates at group boundaries, as illustrated in Figure~\ref{fig:moti2_cache_drop}.
This sudden cache miss occurs because the vector database lacks visibility into query group boundaries, making it unable to prefetch clusters for the next group in advance.
We refer to this limitation as \textit{prefetch invisibility}, which fundamentally restricts the effectiveness of grouping strategies in mitigating cache evictions during group transitions.

\begin{figure}[!t]
    \centering
    \vspace{-8pt}
    \setlength{\tabcolsep}{0.5em} 
    \renewcommand{\arraystretch}{1.0} 
	\begin{tabular}{@{}c@{}c@{}c@{}}
            \includegraphics[width=0.5\linewidth]{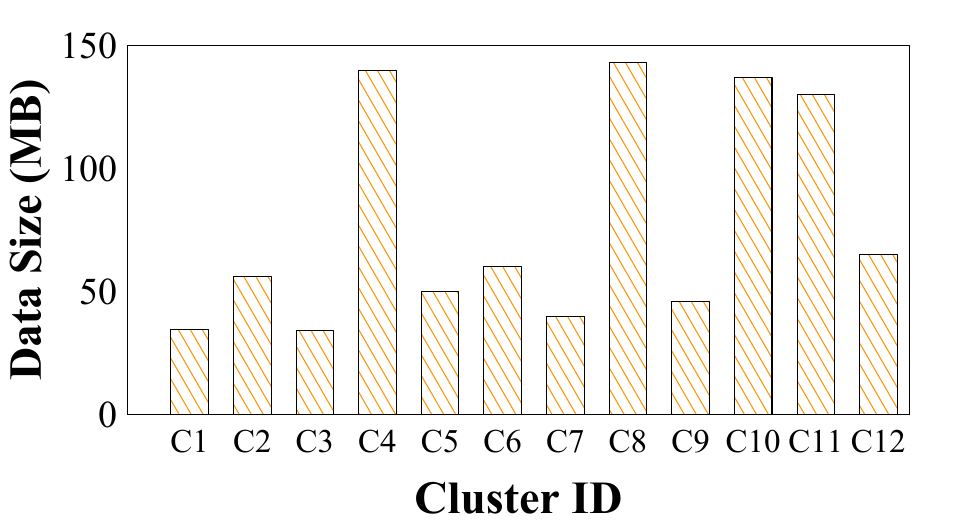} &
        \includegraphics[width=0.5\linewidth, trim={0 0 0 0cm},clip]{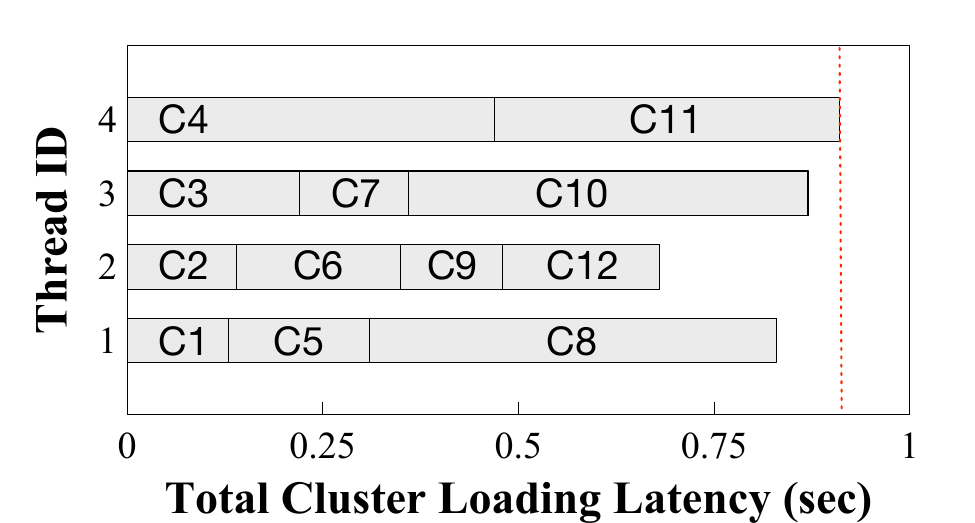} \\
        \small (a) Cluster file size &
        \small (b) Cluster placement per thread \\
    \end{tabular}
        \caption{Cluster size and thread-level loading latency.}
    \label{fig:moti3_cluster_straggler}
    \vspace{-10pt}
\end{figure}

\subsection{Imbalanced Cluster Loading}
To load missing clusters from storage quickly, the system employs multiple cluster loaders that perform parallel disk access.
However, our analysis of FAISS, a widely-used vector database, reveals that it assigns cluster loading tasks to worker threads using round-robin scheduling policy, without accounting for the differences in cluster file sizes.
Assigning cluster loading tasks by cluster ID, without accounting for differences in file sizes, can cause load imbalance across threads.
This scheduling causes load imbalance when cluster sizes vary.
To illustrate this problem, we sorted all 100 clusters by size and selected the top 4 largest and bottom 8 smallest clusters. 
Figure~\ref{fig:moti3_cluster_straggler} (a) shows file size distribution across 12 clusters.
For instance, the largest cluster C10 exceeds 140 MB, while C2 or C9 are under 40 MB.
Figure~\ref{fig:moti3_cluster_straggler} (b) visualizes how these clusters are assigned to threads under round-robin scheduling.
Consequently, Thread 1, assigned the large cluster C10, exhibits a prolonged loading time compared to others. 
While most threads complete their tasks within 0.5 to 0.75 seconds, the entire loading process is stalled until Thread 1 finishes at 0.91 seconds. 
This straggler effect nullifies the benefits of parallelism, as the system must wait for the slowest thread to complete.
This observation indicates that even with well-formed query groups and effective prefetching, performance gains can be significantly undermined if cluster loading is not carefully scheduled to balance workload across threads.
\section{Design}
\label{sec:design}
In this section, we begin by presenting an overview of \proposed{}.
We then provide a detailed explanation of how query grouping is systematically integrated into disk-based vector search to achieve scalable and low-latency when serving vector indexes that exceed host memory capacity.
Additionally, we present a high-level overview of how \proposed{} operates.

\subsection{Overview of \proposed{}}
\vspace{-1pt}

\begin{figure}[!t]
	\centering
    \includegraphics[width=0.8\linewidth]{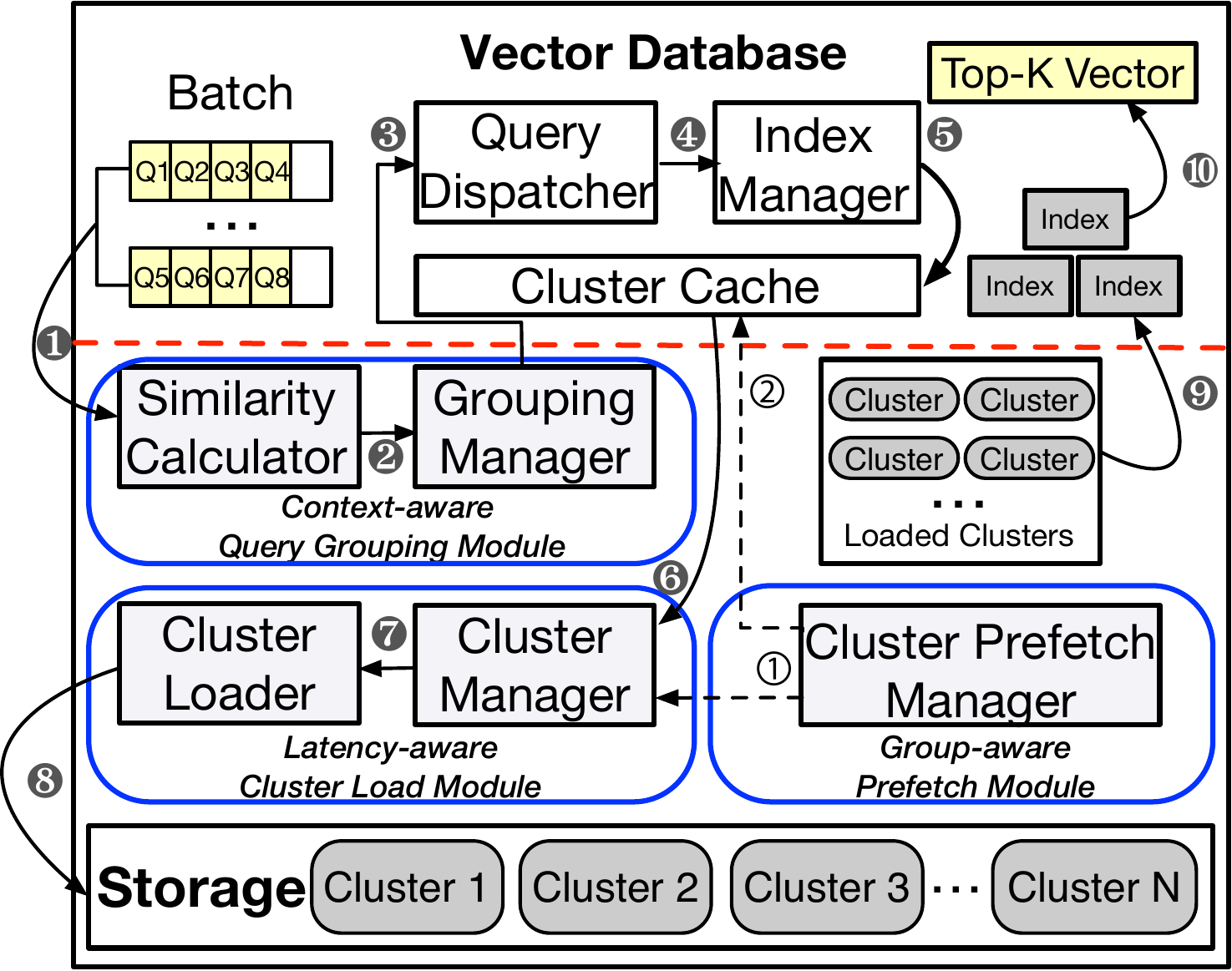}
    \caption{An overview of \proposed{}. It consists of three modules, outlined by \textcolor{blue}{blue} lines. Black arrows indicate synchronous paths, and dotted arrows denote asynchronous ones. Order is marked by numbered circles.}
    \label{fig:overview}
    \vspace{-16pt}
\end{figure}

Figure~\ref{fig:overview} presents an overview of \proposed{}, highlighting its three core modules -- context-aware grouping module, group-aware prefetch module, and latency-aware cluster load module.
\proposed{} is performed after the indexing phase is completed.
As described in Section~\ref{sec:bg_rag}, the embedding model transforms documents into multi-dimensional vectors, which are partitioned into clusters and stored as separate files.
After completing the indexing phase, \proposed{} processes batched queries using three modules.

\begin{figure}[!t]
	\centering
    \includegraphics[width=0.7\linewidth]{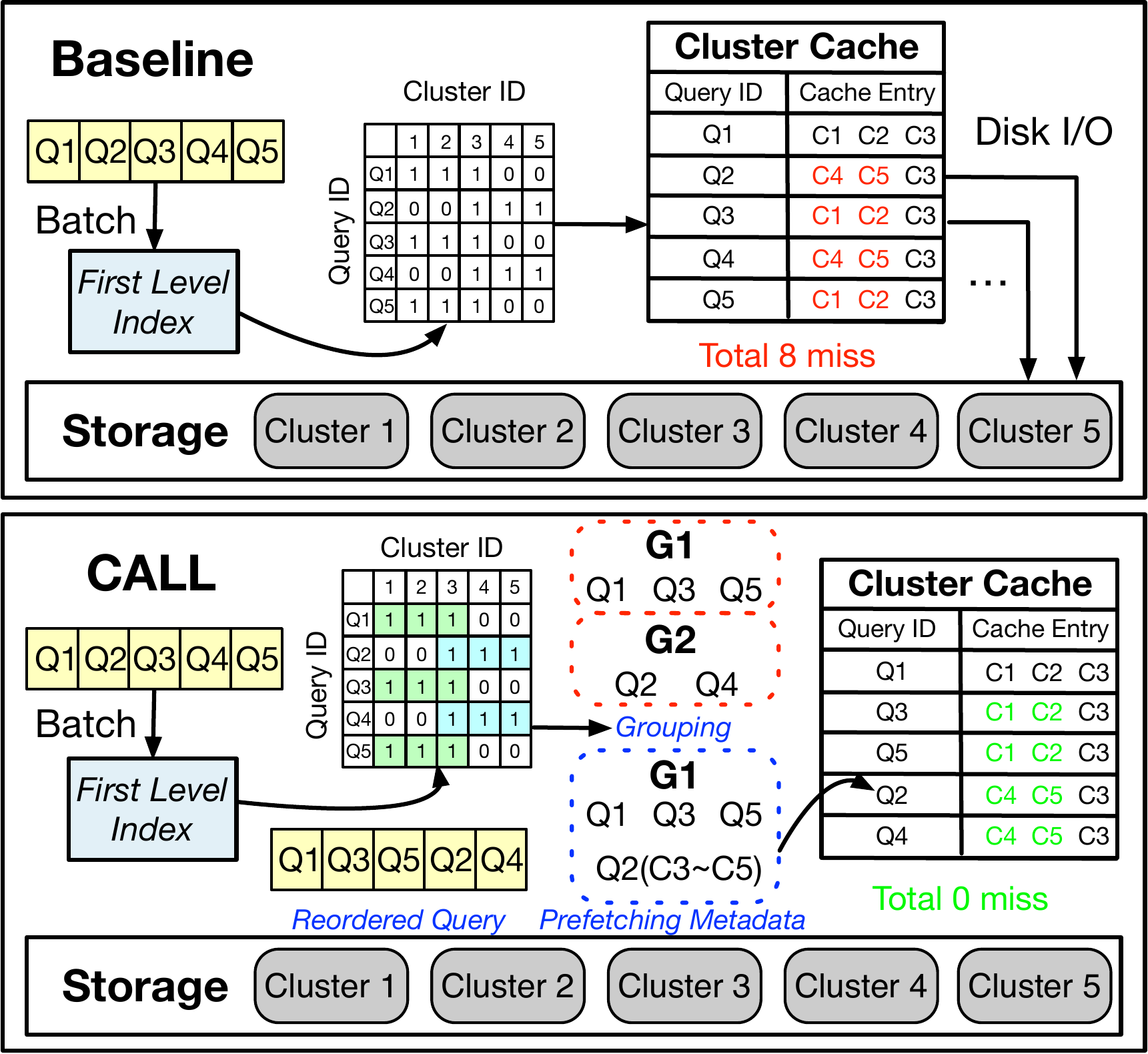}
    \caption{High level scenarios of disk-based vector search for the baseline and \proposed{}. We assume the number of cache entries is 3. Initially, five queries are sent as batch.}
    \label{fig:scenario}
    \vspace{-12pt}
\end{figure}

\noindent\textbf{Context-aware Grouping Module} receives 8 queries (e.g., Q1 $\sim$ Q8) as a batch and probes first level index to identify the set of clusters for each query~\bcircled{\footnotesize{1}}.
Then, \textit{similarity calculator} computes pairwise cluster similarity scores and groups queries with overlapping cluster access patterns~\bcircled{\footnotesize{2}}. 
The queries in the batch are reordered based on similarity scores and forwarded to vector database~\bcircled{\footnotesize{3}}.
The reordering of queries within a batch is described in Figure~\ref{fig:scenario}.
Upon receiving the queries, query dispatcher sends them to \textit{index manager}~\bcircled{\footnotesize{4}}.
It first attempts to load required clusters from \textit{cluster cache cache}~\bcircled{\footnotesize{5}}.
For missing clusters, \textbf{Latency-aware Cluster Load Module} is invoked to fetch candidate clusters from storage~\bcircled{\footnotesize{6}}. 
Subsequently, \textit{cluster manager} provides metadata including cluster size, storage path to \textit{cluster loader}~\bcircled{\footnotesize{7}}, which then performs I/O to read the corresponding files from the storage~\bcircled{\footnotesize{8}}.
Hereafter, partial indexes are constructed using both cached and clusters from the storage~\bcircled{\footnotesize{9}}.
Finally, top-k search is performed across the partial indexes, after which results are sorted and the k-nearest vectors are returned~\bcircled{\scriptsize{10}}.
\noindent\textbf{Group-aware Prefetch Module} monitors the execution order within each query group.
Upon completion of the last query in the current group, \textit{cluster prefetch manager} proactively loads clusters~\wcircled{\footnotesize{1}} required by the first query in the subsequent group into the cache~\wcircled{\footnotesize{2}}.
\vspace{-0.1in}

\subsection{Context-aware Grouping Module}
\label{subsec:context_qg}
To determine cluster similarity across the queries, we adopted Jaccard index.
Cluster similarity calculator computes the similarity with the proportion of the size of the intersection and the union of two sets.
For example, a query stream is defined as $Q = \{ q_1, q_2, \dots, q_n \}$, where each query $q_i$ 
has a set of clusters $C(q_i)$.
The Jaccard similarity is then computed using their respective cluster sets, as shown in Equation~\ref{equation2}.


\vspace{-10pt}
\begin{equation}
J(q_i, q_j) = \frac{|C(q_i) \cap C(q_j)|}{|C(q_i) \cup C(q_j)|}, \text{} C(q_j) = \{ c_1, c_2, \dots, c_p \}
\label{equation2}
\end{equation}

\vspace{-14pt}
\begin{equation}
G_k = \{ q_i \in Q \mid J(q_i, q_j) \geq \theta, \forall q_j \in G_k \}
\label{equation3}
\end{equation}

Based on the computed similarity scores, we applied agglomerative clustering, a hierarchical clustering method that groups similar items based on pairwise distances~\cite{aggogrouping} with a predefined Jaccard distance threshold $\theta$ to form query groups using Equation~\ref{equation3}.
If the similarity score is within the distance threshold $\theta$, the query is assigned to the corresponding group.
Otherwise, a new query group is created, and the query is assigned accordingly.
By applying the threshold-based similarity scoring to all queries, group manager constructs a set of query groups $G = \{ G_1, G_2, \dots, G_n \}$.
\proposed{} performs grouping once for each incoming batch of queries. 
While the current batch is being executed, newly arriving queries are accumulated and processed as the next batch.

Due to variability in user activity and external events, query arrival rates often exhibit sudden bursts.
Hence, it is critical not only to improve cache efficiency via query grouping but also to ensure that the grouping process scales well with traffic and incurs minimal overhead.

Naive Jaccard index computation compare similarities between cluster sets $C(q_i)$ and $C(q_j)$ for all query pairs, where each cluster set is stored as a hash table. 
Although hash tables offer offer average \( O(1) \) lookup time, they rely heavily on key comparisons and conditional branching to verify an overlap of clusters.
During Jaccard similarity computation, repeated set membership checks (e.g., if cluster in set) introduce numerous conditional branches.
Since the access patterns across cluster sets are irregular and input-dependent, these branches are hard to predict, resulting in frequent branch mispredictions.
To avoid the branch-heavy control flow, we introduce a vectorized Jaccard computation method.
\vspace{-10pt}

\begin{equation}
\mathbf{v}_i = [b_1, b_2, \ldots, b_K], \quad
b_k =
\begin{cases}
1 & \text{if } c_k \in C(q_i) \\
0 & \text{otherwise}
\end{cases}
\label{equation4}
\vspace{-8pt}
\end{equation}

\begin{equation}
\mathbf{I}_{i,j} = \mathbf{v}_i \cdot \mathbf{v}_j^\top, \quad
|\mathbf{v}_i \cup \mathbf{v}_j| = \|\mathbf{v}_i\|_1 + \|\mathbf{v}_j\|_1 - \mathbf{I}_{i,j}
\label{equation5}
\vspace{-8pt}
\end{equation}

\begin{equation}
J(q_i, q_j) = \frac{\mathbf{I}_{i,j}}{|\mathbf{v}_i \cup \mathbf{v}_j|}
\label{equation6}
\end{equation}

For $Q_i$, its vectorized cluster set is represented as Equation~\ref{equation4}.
This maps the entire cluster space into a contiguous memory block.
Through a vectorized data structure, pairwise intersections and union cardinalities can be computed using matrix-based bitwise operations as shown in Equation~\ref{equation5}.
This eliminates conditional branching by replacing set membership checks with SIMD-friendly instructions.
As a result, the Jaccard similarity can be computed efficiently without control-flow divergence, as shown in Equation~\ref{equation6}. 
Its branch-less formulation significantly accelerates query grouping.

\subsection{Group-aware Prefetch Module}
\label{subsec:group_prefetch}
Query grouping is performed once per batch, and the assigned groups remain fixed throughout processing. However, despite this grouping, the query dispatcher cannot detect where the group boundaries lie within a batch.
This limitation can delay cluster loading during group transitions, leading to cache misses.
To overcome this limitation, we introduce a lightweight prefetching metadata $P$, defined in Equation~\ref{equation:prefetch_dict}.
Here, $FQ$ denotes the first query of the next group $G_{i+1}$, while $FQSET$ represents the set of clusters of $FQ$.
This metadata is attached to each group during the grouping phase and forwarded to the query dispatcher module.
\vspace{-3pt}
\begin{equation}
\small
P = \{ \text{FQ}: Q_{i+1}(q_f),\ \text{FQSET}: C(Q_{i+1}(q_f)) \}
\label{equation:prefetch_dict}
\end{equation}

Figure~\ref{fig:scenario} illustrates the vector search steps in baseline and \proposed{}.
We assume that cluster cache loads cluster files of first query Q1.
Since the baseline does not account for cluster similarities across the queries, it searches them in sequential order.
Before building the vector index, it checks the cache, but because each query accesses different clusters, a sequence of cache misses occurs. This results in eight consecutive cache misses, each triggering disk I/O during the search.

In contrast, \proposed{} performs grouping on batched queries and generates metadata that encodes group information, including the cluster IDs required by the first query of the subsequent group.
Based on the grouping, the queries within the batch are reordered to maximize cache hit ratios, resulting in consecutive cache hits. 
Specifically, \proposed{} asynchronously preloads cluster files (e.g., C4, C5 for Q2) immediately after completing the vector search for Q5, using the prefetching metadata.
As a result, all required clusters are already cached at query time, leading to zero cache misses.

\vspace{-5pt}
\begin{algorithm}
\footnotesize
\begin{flushleft}
\hspace*{\algorithmicindent} \textbf{Input :} $M = \{m_1, m_2, ..., m_N\}$, $\text{Size}(c_i)$, thread count $T$ 
\hspace*{\algorithmicindent} \textbf{Output :} Load order $L = [c_{i_1}, c_{i_2}, ..., c_{i_N}]$
\end{flushleft}
\begin{algorithmic}[1]
\Procedure{LoadWeightedGreedyPacking}{}
    \State Sort $M$ in descending order by $\text{Size}(m_i)$
    \State Initialize empty thread groups $G_1, \dots, G_{\lceil N/T \rceil}$
    \For{each cluster $m$ in sorted $M$}
        \State Find the earliest thread group $G_j$ such that $|G_j| < T$
        \State Assign $m$ to thread group $G_j$
    \EndFor
    \State Concatenate $L \gets G_1 \Vert G_2 \Vert \dots \Vert G_{\lceil N/T \rceil}$
    \State \Return $L$
\EndProcedure
\end{algorithmic}
\caption{Load Weighted Greedy Packing Algorithm}
\label{alg:gsbs}
\end{algorithm}

\vspace{-0.09in}

\subsection{Latency-aware Cluster Load Module}
\label{subsec:latency_clusterload}
In disk-based vector databases, the size of each cluster file varies significantly due to non-uniform vector, resulting in uneven vector densities across clusters. 
This imbalance becomes problematic when loading multiple clusters in parallel. 
A commonly used scheduling method, such as the round-robin strategy adopted by FAISS, evenly assigns cluster files to threads based on cluster count. 
However, this approach can lead to severe load imbalance because a thread processing a large cluster file becomes a straggler, dominating the total loading time and reducing parallelism efficiency.
To address this straggler problem, we introduce a load weighted greedy packing algorithm that prioritizes the assignment of larger cluster files to earlier thread groups.
As shown in Equation~\ref{eq:cluster_notation}, $M$ denotes the set of missing clusters, where each cluster $m_i$ has an associated size $s_i$.

\vspace{-15pt}
\begin{equation}
\small
M = \{m_1, m_2, \dots, m_N\}, s_i = \text{Size}(m_i), L_j = \sum_{m_i \in G_j} s_i
\label{eq:cluster_notation}
\end{equation}

Then, the total load $L_j$ of thread group $G_j$ is computed as the sum of cluster sizes assigned to that group.
Our goal is to balance $L_j$ across all groups to minimize total cluster loading latency and avoid stragglers.
Algorithm~\ref{alg:gsbs} demonstrates how to achieve our goal to balance the total load $L_j$. 
\section{Evaluation}
\label{sec:evaluation}
\subsection{Experimental Setup}
\label{subsec:experimentalSetup}
\noindent\textbf{Implementation.} 
We implemented \proposed{} based on FAISS version 1.10.0 as vector search engine.
To enable disk-based ANN search, we employed IVF indexing algorithm provided in FAISS library.
During all experiments, we set the total number of clusters to 100, nprobe to 30, and the number of threads to 8.
The cluster cache was configured to hold up to 50 entries, and the Jaccard distance threshold was set to 0.5.
This means that each query creates a partial index using 30 cluster files. Given that the total number of cluster cache entries is 50, up to 50 clusters can reside in the cache.
Upon a cache miss, the cluster cache evicts the least recently used entries first, removing as many entries as the number of missing clusters.
Each experiment included a 1-minute warm-up phase to allow the cache to stabilize.
Our code is available at \textit{\href{https://anonymous.4open.science/r/CALL}{https://anonymous.4open.science/r/CALL}}.

\noindent\textbf{Platform.} 
Our experiments were conducted on a server equipped with Intel(R) i7-8700K 3.70GHz CPU, 16 GB DDR4 DRAM, and a 256GB Samsung 960 NVMe SSD.

\begin{figure}[tb]
    \centering
       \includegraphics[width=0.9\linewidth]{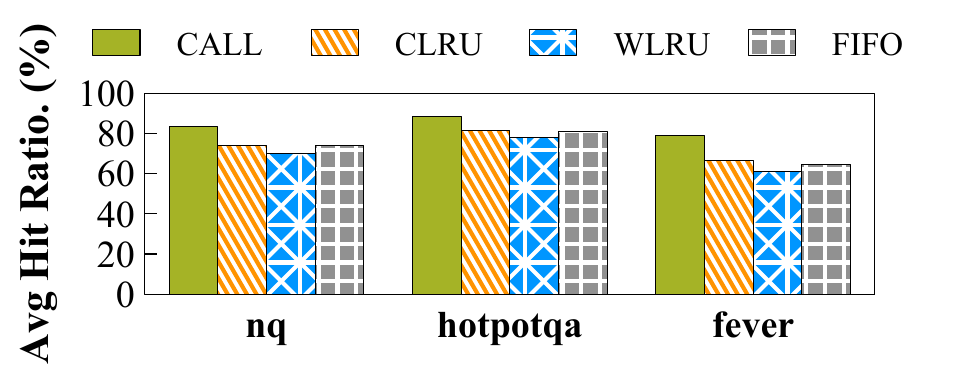}
        \caption{Average cache utilization across datasets under three cache replacement policies.}
    \label{fig:ex_overall_pf_avg_cache_util}
\end{figure}

\begin{figure}[tb]
\centering    
    \begin{tabular}{@{}c@{}c@{}c@{}c@{}}
        \includegraphics[width=0.5\linewidth]{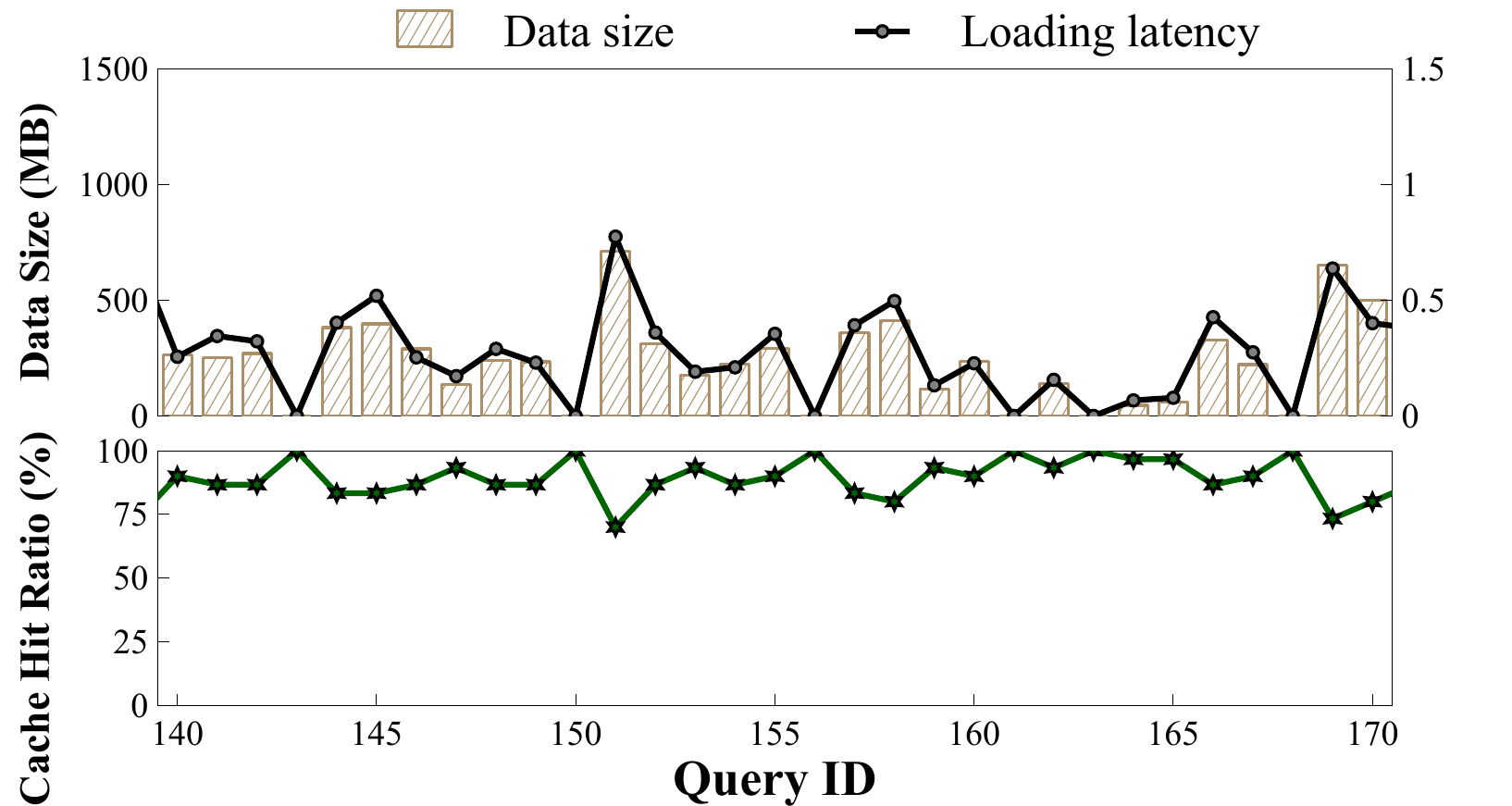} &
        \includegraphics[width=0.5\linewidth]{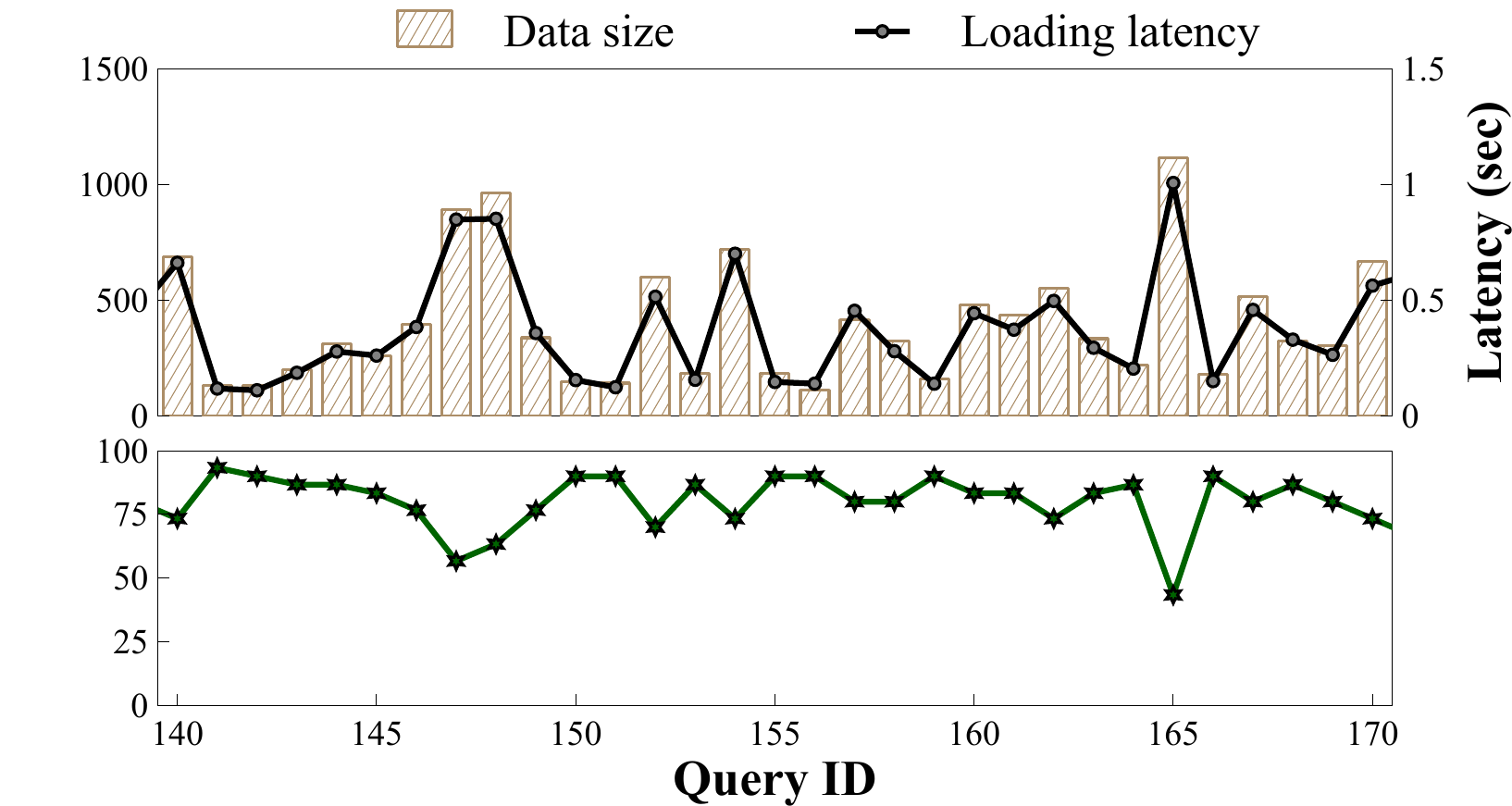} &
        \\ 
        \small (a) \proposed{}{} &
        \small (b) \costlru{}{} &
    \end{tabular} 
        \begin{tabular}{@{}c@{}c@{}c@{}c@{}}
        \includegraphics[width=0.5\linewidth]{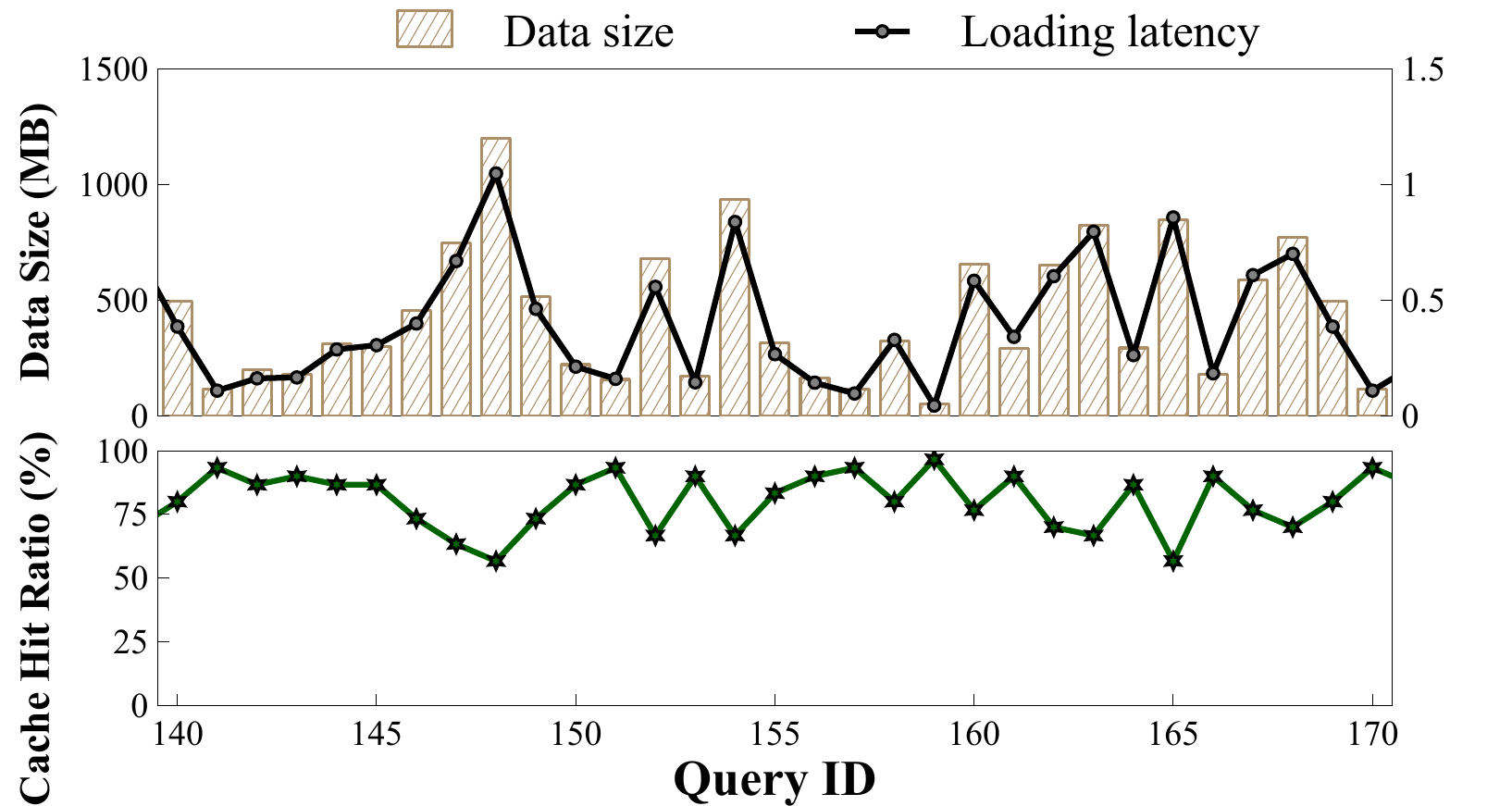} &
        \includegraphics[width=0.5\linewidth]{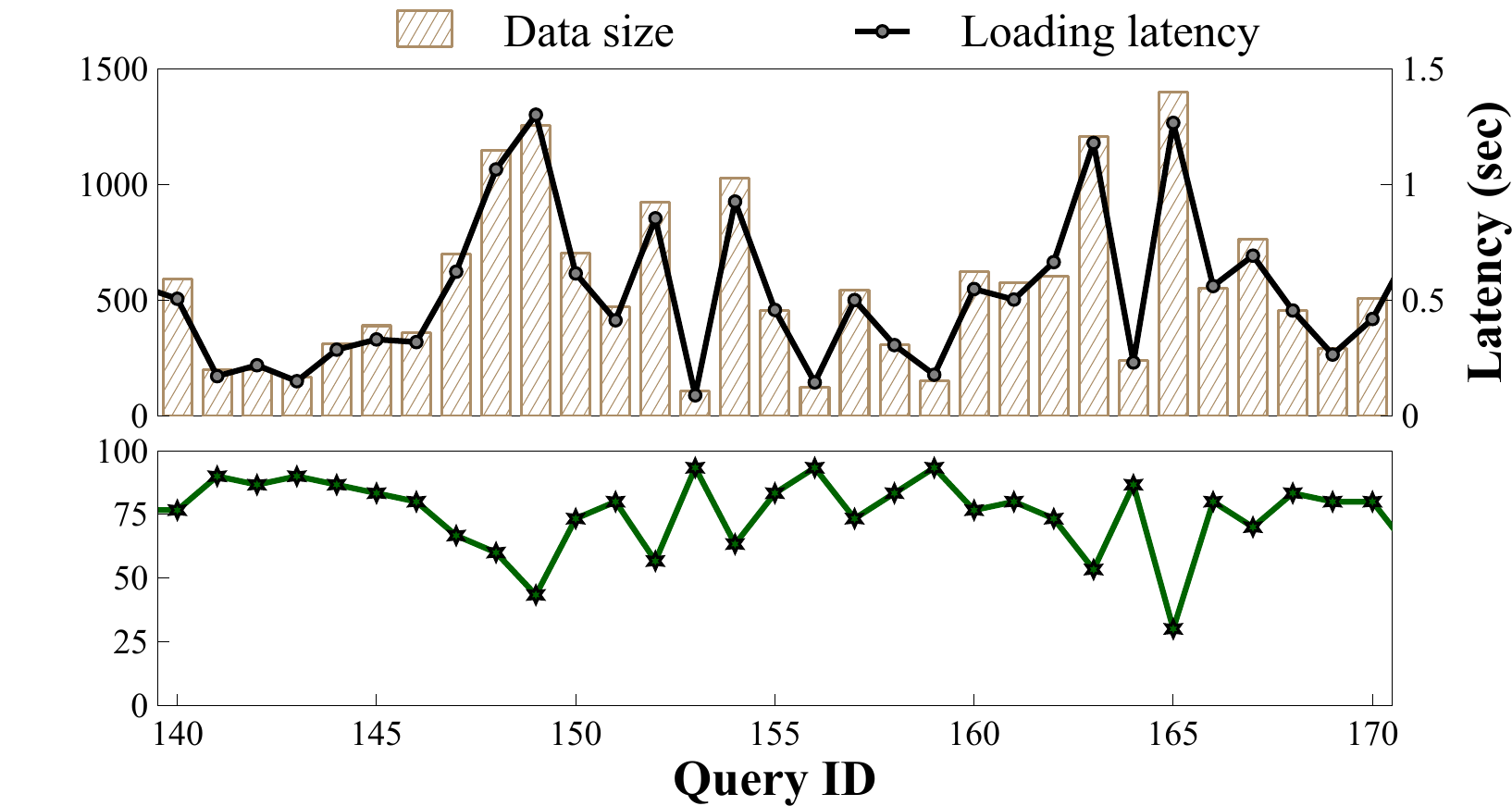} &
        \\ 
        \small (c) \windowlru{}{} &
        \small (d) \fifo{} &
        \end{tabular} 
	\caption{Time-series cache hit ratio, cumulative read size of the clusters, and cluster loading latency under three cache replacement policies.}
	\label{fig:ex_overall_pf_timeseries_cache_util}
    \vspace{-12pt}
\end{figure}


\begin{table}[h]
    \centering 
    \caption{Details of evaluated datasets.}
    \vspace{-5pt}
    \resizebox{1\columnwidth}{!}{    
\begin{tabular}{|c||c|c|c|c|}
\hline
\textbf{Dataset} & \textbf{Records} & \textbf{Embedding Size} & \textbf{Peak Memory Usage} & \textbf{Fit in Memory}\\ \hline\hline
nq~\cite{nq}         & 2.68 M & 8.3 GB  & 14 GB & O \\ \hline
hotpotqa~\cite{hotpotqa} & 5.42 M & 15.4 GB & 25.5 GB & X \\ \hline
fever~\cite{fever}     & 5.23 M & 18.5 GB & 26.4 GB & X \\ \hline
\end{tabular}
    }
    \label{table1:dataset}
\end{table}

\begin{figure*}[!t]
\centering
    \begin{tabular}{@{}c@{}c@{}c@{}c@{}}
        \includegraphics[width=0.33\linewidth]{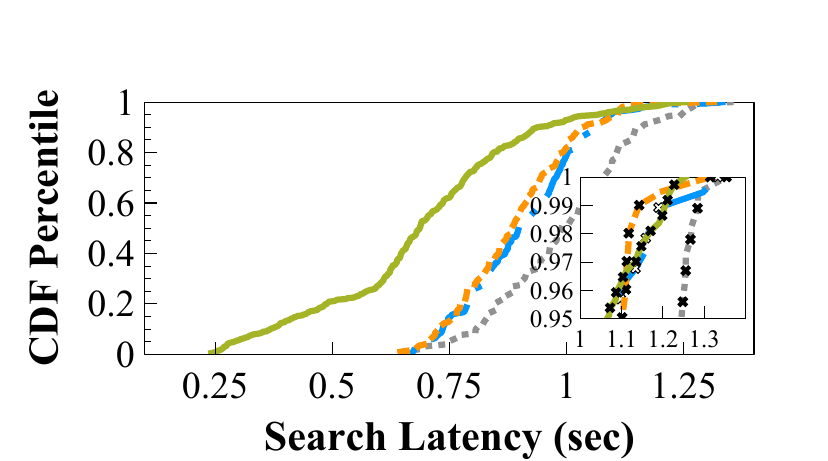} &
        \includegraphics[width=0.33\linewidth]{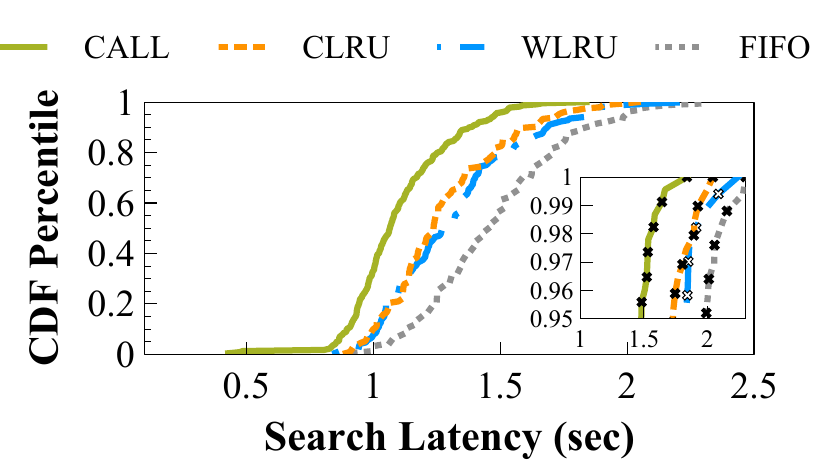} &
        \includegraphics[width=0.33\linewidth]{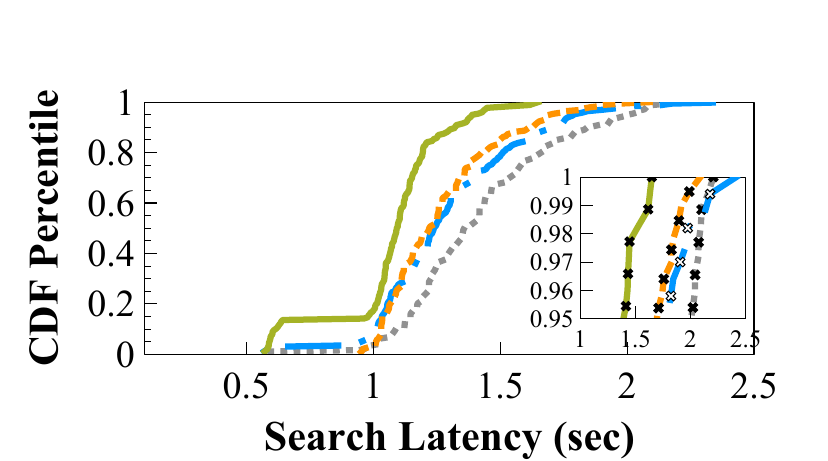} 
        \\
        \small (a) nq &
        \small (b) hotpotqa &
        \small (c) fever \\
    \end{tabular}
	\caption{Tail search latency across datasets under three cache replacement policies.}
	\label{fig:ex_overall_pf_tail}
    \vspace{-10pt}
\end{figure*}

\noindent\textbf{Datasets.}
To construct the vector index, we used three public datasets -- nq~\cite{nq}, hotpotqa~\cite{hotpotqa}, and fever~\cite{fever} from BEIR benchmark~\cite{beir}.
Details of each dataset are provided in Table~\ref{table1:dataset}. 
Despite the total embedding size being smaller than system memory capacity (i.e., nq and hotpotqa), we observed peak memory usage during index construction when loading multiple cluster files concurrently.
This is because each cluster file is first loaded into the kernel page cache and subsequently copied into user space memory.
As a result, both the kernel and user space temporarily hold redundant copies of the same data, doubling memory consumption, thereby exceeding available memory capacity.
Both the corpus and query sets from the same dataset are converted into vector embeddings using all-miniLM-L6-v2 model, which has 384 dimensional dense vector space.

\noindent\textbf{Traffic.}
In real-world scenarios, users often issue multiple queries concurrently, leading to bursty traffic patterns in vector database systems. 
To emulate such behavior, we generate a scalable and bursty query workload using a traffic generator modeled by the Weibull distribution.
During the normal phase, the system issues queries at a steady rate of 100 queries per second. 
At each time interval, there is a 10\% probability of entering a burst phase, during which the query rate surges up to three times the normal rate to simulate sudden traffic surges.

\noindent\textbf{Evaluation Metrics and Comparison Targets.}
We used four evaluation metrics: cache hit ratio, search latency, grouping time, and  cluster loading latency.
The search latency is measured from exploring the cluster cache to vector retrieval.
To evaluate the effectiveness of the proposed design, we implemented three cache replacement schemes and evaluated them with \proposed{}.

\begin{itemize}
    \item \costlru{} prioritizes cache retention based on two metrics that considers both the access frequency of clusters and their associated loading latency\cite{edgerag}.
    \item \windowlru{} retains only the top 10 most frequently accessed clusters within each 1-minute window. 
    \item \fifo{} evicts the cluster that has resided in the cluster cache the longest.
\end{itemize}


\subsection{Overall Performance}
\noindent\textbf{Cache Utilization.}
Figure~\ref{fig:ex_overall_pf_avg_cache_util} shows that \proposed{} consistently achieves higher average cache hit ratios compared to all baselines across the evaluated datasets.
Among them, fever dataset exhibits the most significant gap.
\proposed{} reaches a 92\% hit ratio, while \windowlru{} and \fifo{} fall to 75\% and 60\%, respectively.

To further analyze the enhancement of cache hit ratio, we fixed the dataset fever and measured  time-series cache hit ratio, cumulative read size of clusters, and the cluster loading latency.
As shown in Figure~\ref{fig:ex_overall_pf_timeseries_cache_util}, \proposed{} maintains a stable cache hit ratio throughout the query stream.
As a result, the amount of data read from disk remains low, and the corresponding cluster loading latency consistently stays below 0.5 seconds.
In contrast, \windowlru{} and \fifo{} show sharp drops in cache hit ratio, triggering frequent disk reads and causing loading latency to spike above 1.5 seconds.
This improvement is mainly due to query grouping.
By reordering queries with similar cluster access patterns, clusters are reused efficiently within each group, improving cache utilization.
Additionally, group-aware prefetching helps prevent cache drops during group transitions by loading the next group’s required clusters in advance. 
Even with fewer clusters loaded from disk, \proposed{} avoids stragglers by balancing cluster file sizes across threads, further reducing loading latency.

\noindent\textbf{Tail Latency.}
Figure~\ref{fig:ex_overall_pf_tail} compares the cumulative distribution functions (CDF) of end-to-end search latency across the three datasets.
Similarly, tail latency remains low across all datasets and nearly all time intervals.
In fever dataset, as shown in Figure~\ref{fig:ex_overall_pf_tail} (c), \proposed{} reduces 99th percentile latency by up to 33\%, 32\%, and 21\% compared to \windowlru{}, \fifo{}, and \costlru{}, respectively.
In Figure~\ref{fig:ex_overall_pf_tail} (a), \costlru{} slightly reduces search latency from 95 to 98 percentile.
In addition to \costlru{}, both \windowlru{} and \fifo{} exhibit similar performance at the 96th percentile. 
This is because nq dataset has a relatively small embedding size that fits comfortably in memory, as shown in Table~\ref{table1:dataset}. 
In such settings, cluster reuse is high, and the working set of active clusters remains stable over time.
However, these fail to handle extreme tail cases, such as the 100th percentile.
In the following sections, we analyze the contribution of each module in \proposed{} to reducing search latency.

\begin{table}[!t]
\centering
\caption{Normalized CPU branch miss rate. \proposed{} buffers queries for 3 seconds under low traffic and 10 seconds under high traffic conditions.}
\label{tab:ex_ef_module1_grouping_latency}
\resizebox{\columnwidth}{!}{
\begin{tabular}{ccc|cc}
\toprule
\textbf{Number of}
& \multicolumn{2}{c|}{\textbf{Grouping Time (sec)}} 
& \multicolumn{2}{c}{\textbf{Branch Miss Rate}} \\
\cmidrule(lr){2-3} \cmidrule(lr){4-5}
\textbf{Queries}
& \textbf{Bitmap} & \textbf{Hash Table} 
& \textbf{Bitmap} & \textbf{Hash Table} \\
\midrule
1000(Low) & 0.58 & 3.57 & 1 & 1.27 \\
1500 & 0.85 & 6.46 & 1 & 1.32 \\
2000 & 1.07 & 11.55 & 1 & 1.38 \\
2500(High) & 1.47 & 18.81 & 1 & 1.47 \\
\bottomrule
\end{tabular}
}
\vspace{-12pt}
\end{table}

\begin{figure}[!t]
    \centering
       \includegraphics[width=0.8\linewidth]{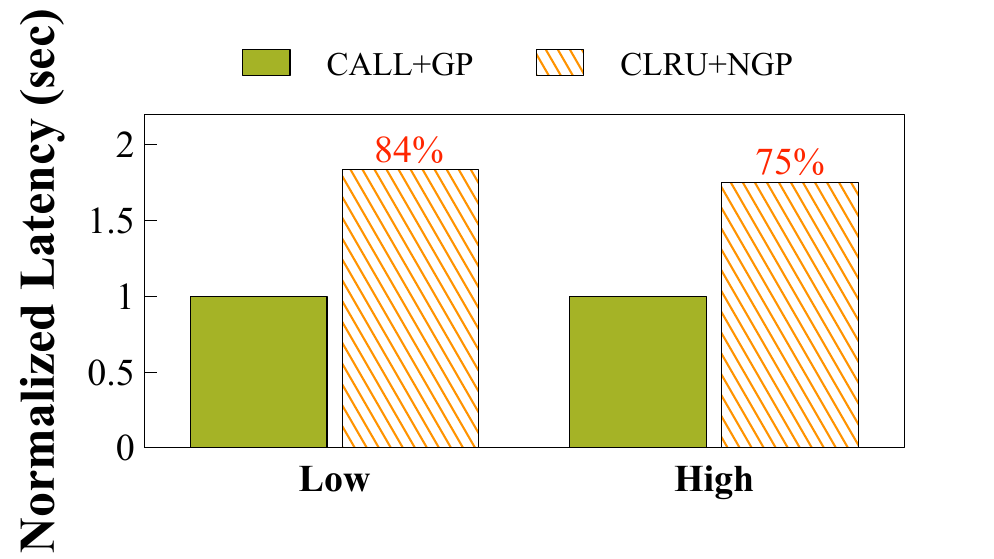}
        \caption{Normalized end-to-end search latency.}
        \vspace{-12pt}
    \label{fig:ex_overhead_total_duration_gp_or_ngp}
\end{figure}

\begin{figure}[!t] 
	\centering
    \includegraphics[width=0.8\linewidth]{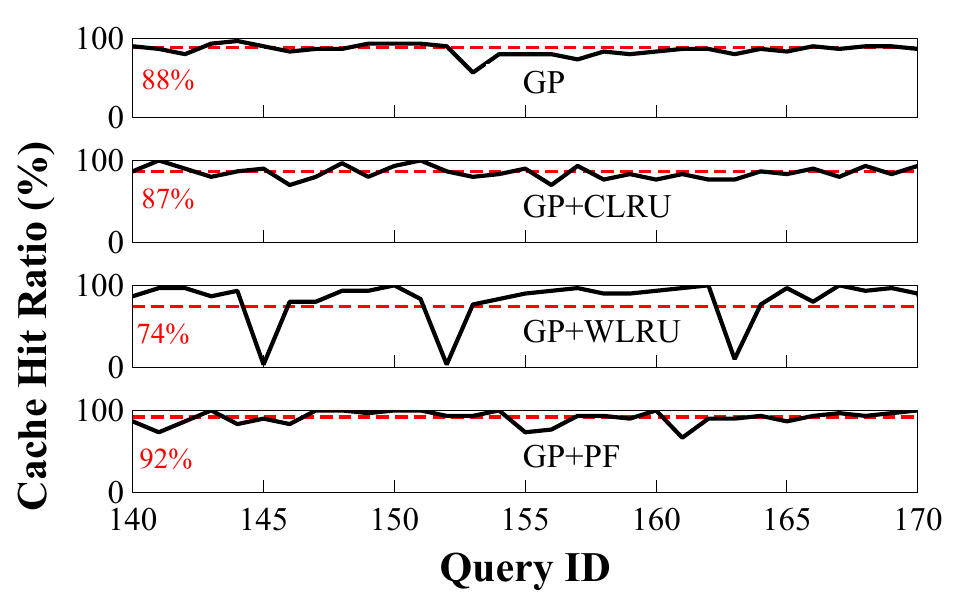}
    \caption{Effectiveness of group-aware prefetch. \textcolor{red}{Red} dotted line indicates an average cache utilization. GP denotes context-aware grouping module, and PF denotes group-aware prefetch module.
    }
    \label{fig:ex_ef_module1_grouping}
    \vspace{-12pt}
\end{figure}



\subsection{Module Effectiveness}
\noindent\textbf{Effectiveness of Context-aware Grouping.}
To ensure that query grouping introduces minimal overhead, we compared the runtime and CPU efficiency of bitmap-based and hash table-based Jaccard computation.
As shown in Table~\ref{tab:ex_ef_module1_grouping_latency}, as the number of queries increases from 1000 to 2500, the grouping time of the bitmap-based method increases slightly, whereas the hash table-based method grows drastically up to 18.81 seconds.
This efficiency stems from the substantially lower branch miss rate achieved by the bitmap-based approach, reducing miss rate by up to 47\%.
As a result, \proposed{} performs query grouping with negligible delay.
Figure~\ref{fig:ex_overhead_total_duration_gp_or_ngp} further confirms this by showing that \proposed{} completes full query processing up to 84\% faster than baseline, even under high traffic.
This improvement stems not only from the scalability of our grouping method, but also from its ability to significantly increase cache hit rates, which directly translates to reduced disk I/O.

\begin{figure}[!t]
    \centering
    \setlength{\tabcolsep}{0.5em} 
    \renewcommand{\arraystretch}{1.0} 
	\begin{tabular}{@{}c@{}c@{}c@{}}
            \includegraphics[width=0.5\linewidth]{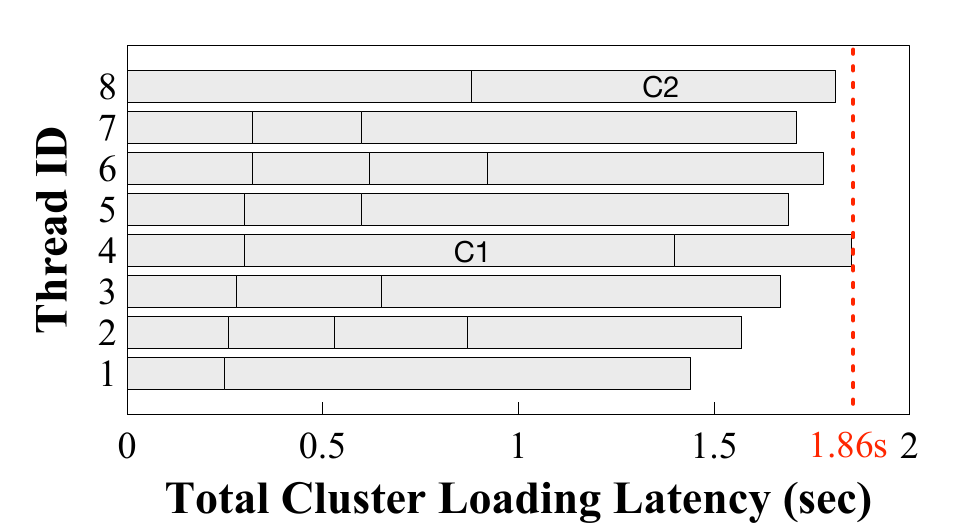} &
        \includegraphics[width=0.5\linewidth]{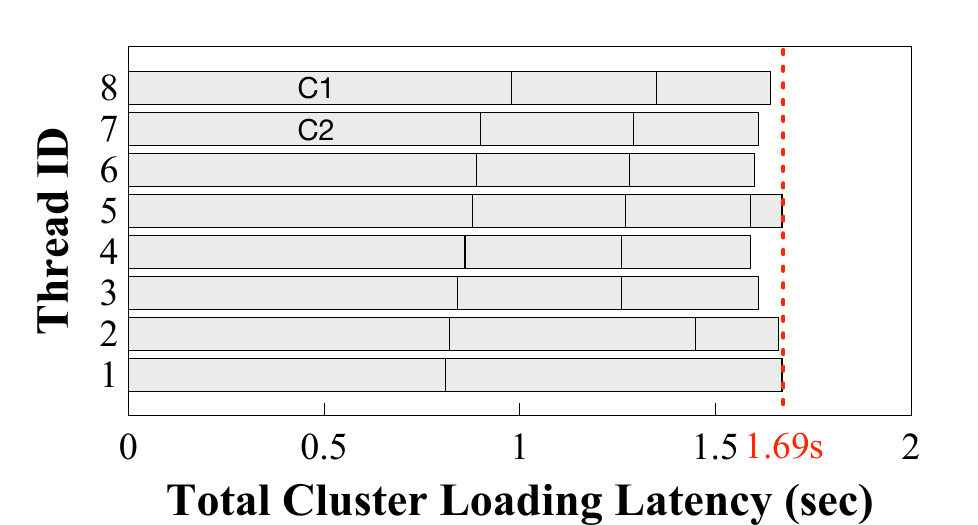} \\
        \small (a) Baseline &
        \small (b) \proposed{} \\
    \end{tabular}
        \caption{Total cluster loading latency across multiple worker threads under varying cluster size distributions. \textcolor{red}{Red} dotted line indicates longest execution time.}
    \label{fig:ex_ef_module3_straggler}
    \vspace{-15pt}
\end{figure}

\noindent\textbf{Effectiveness of Group-aware Prefetch.}
Figure~\ref{fig:ex_ef_module1_grouping} highlights the effectiveness of group-aware prefetching. Even with grouping alone, \proposed{} achieves an average cache hit rate of 88\%, though cache drops still occur at group boundaries. 
By sending prefetching metadata with batched queries, our prefetch module ensures the first query of every group hits the cache, achieving the 100\% cache hit and raising the overall hit rate to 92\%.
In contrast, \costlru{} and \windowlru{}, lacking group awareness, rely solely on frequency-based metrics and fail to prefetch effectively. 
Notably, \windowlru{} occasionally misses all clusters, indicating total prefetch failure.
Notably, \windowlru{} occasionally misses all clusters, indicating a complete prefetch failure.
This issue arises because \windowlru{} preloads clusters that were frequently accessed within a fixed time window (e.g., 1 minute), without considering the actual cluster needs of the incoming query group.
Due to this coarse-grained frequency tracking and lack of group-level visibility, \windowlru{} often prefetches irrelevant clusters, especially at group boundaries.


\noindent\textbf{Effectiveness of Latency-aware Cluster Load.}
Figure~\ref{fig:ex_ef_module3_straggler} illustrates the execution timeline of each thread during parallel cluster loading.
Under the baseline, cluster loading suffers from load imbalance, where thread 4 becomes a straggler due to large cluster C1, leading to a total latency of 1.86 seconds.
In contrast, \proposed{} assigns large clusters across threads using a greedy packing strategy, achieving balanced execution and reducing total cluster loading latency to 1.69 seconds, a 10\% improvement over the baseline.

\vspace{-10pt}
\subsection{Overhead Analysis}
\begin{figure}[!t]
    \centering
    \setlength{\tabcolsep}{0.5em} 
    \renewcommand{\arraystretch}{1.0} 
	\begin{tabular}{@{}c@{}c@{}c@{}}
        \includegraphics[width=0.5\linewidth,trim={0 0 0 0cm},clip]{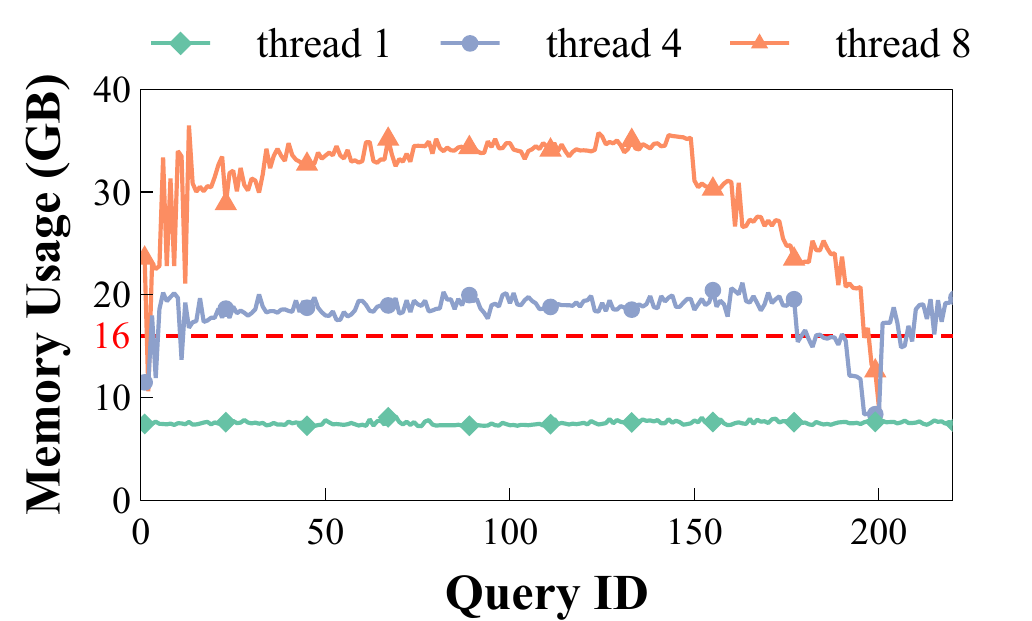} &
        \includegraphics[width=0.4\linewidth,trim={0 0 0 0cm},clip]{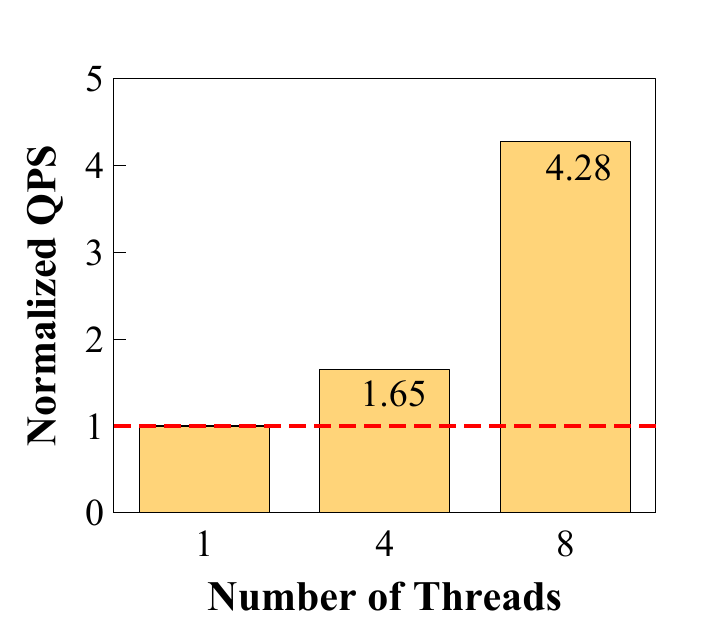} \\
         (a) Memory Usage &
         (b) Normalized QPS \\
    \end{tabular}
        \caption{Impact of the number of search thread on memory usage and query throughput. \textcolor{red}{Red} dotted line in (a) indicates total system memory capacity.}
    \label{fig:thread_mem}
    \vspace{-18pt}
\end{figure}

Figure~\ref{fig:thread_mem} illustrates trade-off between query throughput and memory usage as the number of search threads increases.
While adding threads boosts throughput, which is 1.65 times higher with 4 threads and 4.28 times higher with 8, leading to excessive memory consumption due to redundant cluster loading.
This results in out-of-memory failures when usage exceeds the 16 GB limit.
In contrast, \proposed{} maintains a compact footprint under 8GB by maximizing cache reuse through single-threaded, context-aware grouping.
This design achieves low-latency vector search without memory shortage, making it well-suited for resource-constrained environments.

\vspace{-4pt}
\section{Conclusion}
\label{sec:conclusion}
Through empirical analysis, we reveal that non-uniform cluster access patterns, cache misses during group transitions, and thread-level load imbalance are the key contributors to increasing search latency.
To address this issue, the proposed \proposed{} in this study effectively schedules the queries within a batch by reordering them to maximize cluster cache utilization.
Our extensive evaluation reveals that \proposed{} significantly reduces both average search latency and tail latency compared to existing approaches across various real-worl workloads.


\bibliography{ref}
\footnotesize
\bibliographystyle{ieeetr}
\end{document}